\documentclass[12pt]{article}
\usepackage{amsmath, amsfonts, amsthm, amssymb, graphicx, xcolor}
\usepackage[small]{caption}
\usepackage{colonequals}
\usepackage{geometry}
\usepackage{multicol}
\usepackage{dsdshorthand}
\usepackage{algorithm2e}
\usepackage{color}
\usepackage{footmisc}
\definecolor{darkblue}{rgb}{0.1,0.1,.7}
\usepackage[colorlinks, linkcolor=darkblue, citecolor=darkblue, linktocpage]{hyperref}
\usepackage[square, comma, sort&compress,numbers]{natbib}
\SetKwRepeat{Do}{do}{until} 
\RestyleAlgo{ruled}
\usepackage{subcaption}
\usepackage{authblk}

\renewcommand{\d}{\delta}

\newcommand{\bigO}{\mathcal O}

\makeatletter
\def\@fpheader{\ }
\makeatother

\title{\Large {\bf Skydiving to Bootstrap Islands}}
\author[a]{Aike Liu}
\author[a]{David Simmons-Duffin}
\author[b]{Ning Su}
\author[c]{Balt C. van Rees}
\affil[a]{Walter Burke Institute for Theoretical Physics, Caltech, Pasadena, California 91125, USA}
\affil[b]{Dipartimento di Fisica E. Fermi, Universit\`a di Pisa Largo Bruno Pontecorvo 3, I-56127 Pisa, Italy}
\affil[c]{CPHT, CNRS, \'Ecole polytechnique, Institut Polytechnique de Paris, 91120 Palaiseau, France}

\begin{document}

\begin{titlepage}

\maketitle
\thispagestyle{empty}

\begin{abstract}
We study families of semidefinite programs (SDPs) that depend nonlinearly on a small number of ``external" parameters. Such families appear universally in numerical bootstrap computations. The traditional method for finding an optimal point in parameter space works by first solving an SDP with fixed external parameters, then moving to a new point in parameter space and repeating the process. Instead, we unify solving the SDP and moving in parameter space in a single algorithm that we call ``skydiving". We test skydiving on some representative problems in the conformal bootstrap, finding significant speedups compared to traditional methods.

\end{abstract}
\end{titlepage}

\newpage

\setcounter{page}{1}

\tableofcontents

\newpage

\section{Introduction}

The numerical conformal bootstrap (see \cite{Poland_2019, poland2022snowmass} for recent reviews) has revealed that small sets of crossing equations can encode detailed information about a CFT. But how does this information scale as we study more crossing equations? Unfortunately, larger systems of crossing equations pose new numerical challenges. More equations depend on more parameters --- for example the scaling dimensions and OPE coefficients of ``external" operators entering the given correlators.\footnote{We refer to an operator as ``external" if it appears in a four-point correlator in a given bootstrap problem. For example, when bootstrapping $\<\cO_1\cO_2\cO_3\cO_4\>$, the operators $\cO_1,\cO_2,\cO_3,\cO_4$ are all ``external." We refer to their scaling dimensions and the OPE coefficients among them as ``external" dimensions and OPE coefficients. By contrast, an ``internal" operator is one that appears in OPEs of the external operators, but is not itself external.} In order to build exclusion plots in the space of CFT data, we must find efficient ways to search over these parameters. 

The navigator function of \cite{Reehorst:2021ykw} is a useful tool for this problem. It provides a notion of ``height" on the space of CFT data. By following the navigator function to its minima, we can find allowed points in this space. In this work, we propose an efficient method for using the navigator function without computing it in detail. Our approach is  analogous to skydiving onto a landscape instead of walking across it.

Let us recall the idea behind the navigator function in more detail. Numerical bootstrap problems depend on a parameter space $\cP$ that can include external scaling dimensions, gap assumptions, OPE coefficients, etc.. There exists a binary function on $\cP$, indicating whether or not a point $p \in \cP$ is consistent with crossing symmetry. In practice, this function is computed by solving a semidefinite program (SDP) \cite{Poland:2011ey,Kos:2014bka,Simmons-Duffin:2015qma}. If a so-called ``dual" solution of the SDP exists, the point is inconsistent with crossing symmetry, and we deem it {\it infeasible}. If no ``dual" solution exists, the point may be consistent with crossing symmetry, and we deem it {\it feasible}. The ultimate goal is to carve away all the infeasible points until only a small feasible region remains, thus determining the possible values of the physical parameters with an accuracy given by the size of the feasible region. 

The navigator function of \cite{Reehorst:2021ykw} replaces this binary function ({\it feasible} vs.\ {\it infeasible}) with a real-valued function that is positive for infeasible points and negative for feasible points. Crucially, the navigator function is continuous and locally differentiable, and  its gradient provides information about where to navigate in $\cP$. For example, a simple numerical minimization algorithm can be used to navigate towards a feasible point. It is also easy to find a feasible point with, say, the largest value of a given coordinate in $\cP$. As discussed in \cite{Reehorst:2021ykw}, this is a straightforward equality-constrained optimization problem.

The searches discussed in \cite{Reehorst:2021ykw} involve two steps for each point $p\in \cP$. First, one computes the navigator function $\cN(p)$ and its gradient $\nabla \cN(p)$ by solving an SDP to high precision. Then, one uses these values to determine a step $p \to p + \delta p$ in $\cP$. If we denote the internal variables of the SDP solver as $\xi$, this two-step process changes $\xi$ and $p$ in an alternating fashion. This is suboptimal, in particular because the optimal $\xi$ variables are solved from scratch for each $p$ (and the usual SDP algorithms do not allow for efficient warm-starting from a previous solution). Instead, it is much more natural to update both $p$ and $\xi$ simultaneously at every step, which is what we do in this work.

\begin{figure}[!ht]
	\centering
	\includegraphics[width=0.8\textwidth]{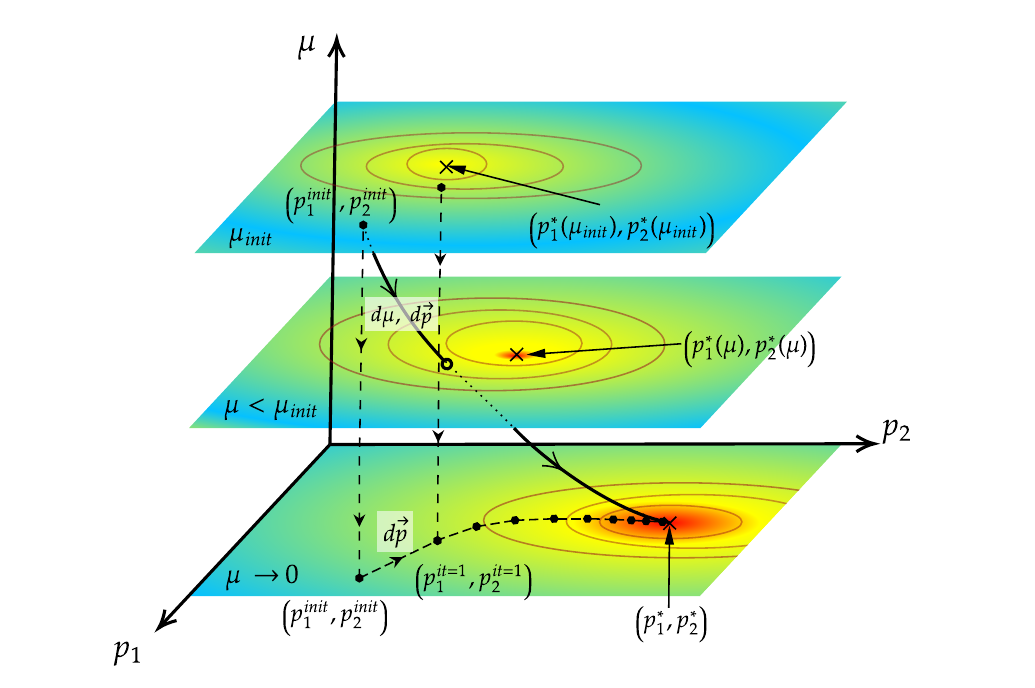}
	\caption[]{\label{fig:compare_with_navigator} A comparison of (1) repeatedly computing the navigator function vs.\ (2) skydiving. Here, the variable $\mu$ measures the distance to a solution of an SDP. The SDP is solved when $\mu\to 0$. The vector $p=(p_1,p_2)$ represents external parameters. The traditional navigator approach solves an SDP for each $p$, beginning at large $\mu$ and converging along the dashed vertical lines towards $\mu=0$ each time. Meanwhile, skydiving moves in $p$ and $\mu$ simultaneously, following the solid black curve.}
\end{figure}

Beginning with initial values of the solver variables $\xi$ and external parameters $p$, we compute a simultaneous step in both variables, and follow it to a new value of $\xi$ and $p$. As the optimization continues, we converge simultaneously to an optimal value of $\xi$ that solves the SDP and the desired value of $p$ that optimizes some user-defined criterion. Thus, finding an optimal point in $\cP$ space occurs over a comparable time scale to the solution of a {\it single} SDP. This is a substantial speedup over traditional approaches, which typically require solving tens, hundreds, or even thousands of auxiliary SDPs, see figure~\ref{fig:compare_with_navigator}.

For fixed $p$, the space of allowed $\xi$ is convex, and thus $\xi$ can be optimized deterministically using interior point methods. However, once we allow $p$ to vary, our optimization problem is no longer convex (though violations of convexity occur in a relatively small number of dimensions).  We therefore find ourselves in the less-familiar world of non-convex constrained optimization, or more precisely non-linear semidefinite programming. In this work, we focus on families of SDPs that arise in typical conformal bootstrap problems --- in particular families of Polynomial Matrix Problems (PMPs) \cite{Simmons-Duffin:2015qma} depending on a small number of parameters. We find solution methods for these problems that work effectively in many examples, including ways to mitigate issues like stalling described below. 
Although our strategy shows promising results in the examples we have considered, we cannot claim it to be optimal or universally applicable. Continuing to explore and develop other strategies remains an interesting problem for future research.

This paper is organized as follows. In section~\ref{sec:sdp}, we review the framework of semidefinite programming and explain how to incorporate external parameters.  In section~\ref{section:Skydiving_Algorithm}, we introduce the skydiving algorithm for solving families of SDPs. We discuss details of our implementation of skydiving in section~\ref{sec:Implementation}. In section~\ref{sec:examples}, we present examples of applying skydiving to various bootstrap problems, including the 3d Ising model and 3d O(3) model, comparing with previous methods. Finally, we conclude in section~\ref{sec:conclusion} with a discussion of some current limitations and potential improvements to skydiving, and possible future applications.

\section{Non-linear semidefinite programming problems} \label{sec:sdp}

\subsection{Review of semidefinite programming}

\label{subsec:review}
\subsubsection{Problem formulation}
Using the methods of \cite{Poland:2011ey,Kos:2014bka}, a numerical conformal bootstrap problem can be formulated as a semidefinite program (SDP). Let us briefly review the solution procedure used by the semidefinite program solver \texttt{SDPB} \cite{Simmons-Duffin:2015qma,Landry:2019qug}. Following the notation of \cite{Simmons-Duffin:2015qma}, an SDP takes the form:
\begin{equation}
\begin{split}
\text{maximize  } &b^T y  \text{ over } y \in \mathbb R^n, Y \in \mathcal S^K\\
\text{\qquad such that  } &Y \succeq 0\quad\text{and}\\
 &B y + \Tr(A_* Y) = c\,
\label{DB}
\end{split} \qquad\qquad (\textrm{dual}),
\end{equation}
where $\mathcal S^K$ denotes the space of symmetric matrices of size $K$. Here $c \in \mathbb R^P$ is a vector, $B \in \mathbb (\mathbb R^{n})^P$ is a rectangular matrix, and $A_*=(A_1,\ldots,A_P) \in  (\mathcal S^K)^P$ is a vector of matrices. The program \eqref{DB} is called the \emph{dual} program. The corresponding \emph{primal} program is:
\begin{equation}
\begin{split}
\text{minimize  } & c^T x \quad \text{over} \quad x \in \mathbb R^P   \\
\text{\qquad such that  } & X(x) \colonequals x^T A_* \succeq 0 \quad\text{and}\\
 & B^T x = b\,
\label{PB}
\end{split}
\qquad\qquad (\textrm{primal}),
\end{equation}
where $x^T A_*\equiv \sum_{p=1}^P x_p A_p \in \mathcal S^K$.

A vector $x$ is called \emph{primal feasible} if the conditions in \eqref{PB} are satisfied, even if optimality is not necessarily achieved. Similarly, a pair $(y, Y)$ is \emph{dual feasible} if it obeys the conditions in \eqref{DB}, without necessarily achieving optimality. The \emph{duality gap} is the difference between the primal and dual objectives:
\begin{equation}
D(x, y) \colonequals c^T x - b^T y\,.
\end{equation}
Positive semidefiniteness of $X$ and $Y$ implies that the duality gap is nonnegative when $x$ is primal feasible and $(y,Y)$ is dual feasible:
\begin{equation}
D(x, y) = \Tr( (x^T A_*)\,Y) = \Tr(X Y) \geq 0\,,
\end{equation}

Let us denote the values at optimality of the primal and dual problems by $(x_*,X_*)$ and $(y_*,Y_*)$, respectively. It is a standard result that, for a generic SDP, the duality gap \emph{vanishes}:
\begin{equation}
	c^T x_* = b^T y_*
\end{equation}
This, in turn, implies the \emph{complementarity condition}:
\begin{equation}
\label{oldcomplementarity}
	X_* Y_* = 0,
\end{equation}
which follows from $\Tr(X_* Y_*) = 0$ and positive semidefiniteness of $X_*$ and $Y_*$.

\subsubsection{Solution algorithm}
In \texttt{SDPB}, SDPs are solved using a \emph{primal-dual interior point algorithm}. This algorithm solves both the primal and dual problems simultaneously and approaches optimality from the interior of the cone of positive semidefinite matrices, meaning $X, Y \succ 0$ at every step. An efficient way to describe the algorithm is via the Lagrange function
\begin{equation}
    L(x,y,X,Y) = c^T x + b^T y - x^T B y + \Tr((X - x^T A_*)Y) - \mu \log \det X,
    \label{LF}
\end{equation}
where $\mu$ is a parameter multiplying a logarithmic \emph{barrier function} $-\log \det X$ that is meant to help $X$ stay within the cone of positive semidefinite matrices. The stationarity equations of this Lagrange function with respect to $x$, $y$, and $Y$ yield the primal and dual feasibility conditions, while stationarity with respect to $X$ yields:
\begin{equation}
\label{newcomplementarity}
    X Y = \mu I,
\end{equation}
which is a finite-$\mu$ version of the complementarity condition \eqref{oldcomplementarity}.

The \emph{central path} is the $\mu$-dependent line of stationary points of the Lagrange function \eqref{LF}. The solution to the original SDP is obtained by taking $\mu \downarrow 0$ along the central path. A standard Newton step $(dx,dX,dy,dY)$ towards this central path can be computed by solving the linearized equations:\footnote{As explained in \cite{Simmons-Duffin:2015qma}, the natural way to solve these equations leads to an update $dY$ that is not manifestly symmetric. In practice $Y$ is therefore updated with the symmetrized version of the $dY$ computed from the Newton step equations. This will be left implicit in our notation.} 

\be
\label{newtonstep}
	dx^T A_* - dX &= x^T A_* - X\nn\\
	B^T dx &= b - B^T x\nn\\
	B dy + \Tr(A_* dY) &= c - B y - \Tr(A_* Y)\nn\\
	dX\, Y + X dY &= R \colonequals \mu I - X Y.
\ee
These are solved in practice with the following procedure. First one solves the linear system:
\begin{equation}
\begin{pmatrix}
S & B\\
B^T & 0
\end{pmatrix}
\begin{pmatrix}
x + dx \\
y + dy
\end{pmatrix}
= 
\begin{pmatrix}
B y - c + \Tr(A_* Y) + \mu \Tr(A_* X^{-1})\\
b
\end{pmatrix},
\end{equation}
and then one can compute the updates:
\be
	dX &= (x^T  + dx^T) A_* - X\nn\\
	dY &= X^{-1} (R - dX Y).
\ee
Here the matrix $S$ is defined element-wise as $S_{ij} = \Tr(A_i X^{-1} A_j Y)$. It is easily verified that this procedure is a simple rewriting of the system \eqref{newtonstep}. One important problem is however that this linear system (generically) becomes singular at optimality. First of all, $X$ and $Y$ become singular because $XY = 0$ and $X,Y \succeq 0$, and $S$ becomes singular because $x^T S x = \Tr(XY) = 0$. In practice, the singularity of the linear system \emph{at} optimality can cause serious numerical instabilities \emph{near} optimality.

The finite-$\mu$ version of the problem has the advantage that $X$ and $Y$ are (again, generically) strictly positive definite even at optimality, since they now obey equation \eqref{newcomplementarity}. This makes the Newton steps towards the modified complementarity condition \eqref{newcomplementarity} well-defined. But of course we would be solving the wrong problem if we kept $\mu$ fixed throughout the run. In practice one therefore gradually decreases $\mu$. At every step one estimates a `current' value of $\mu$ as
\begin{equation}
	\mu_\text{est} = \Tr(X Y) / \dim(X),
\end{equation}
and then one takes the above Newton step with
\begin{equation}
\label{betadefinition}
	\mu \leftarrow \beta \mu_\text{est},
\end{equation}
with $\beta \in (0,1)$ chosen by the user; a typical value is 0.3. In this way, one obtains a trajectory where each step is aimed towards an ever lower point along the central path. 

The Newton step itself is not guaranteed to maintain positivity of the matrices $X$ and $Y$. In practice one therefore chooses an update
\be\label{eq:SDPB_updates}
(X,x) &\leftarrow (X,x) + \alpha_P (dX,dx),\nn\\
(Y,y) &\leftarrow (Y,y) + \alpha_D (dY,dy),
\ee
where $\alpha_P$ and $\alpha_D$ are chosen in $(0,1]$ to ensure that $X$ and $Y$ (and therefore also $S$) remain inside the cone of positive semidefinite matrices.

The essential balance of this algorithm arises from the desire to decrease $\mu$ quickly while maintaining reasonable $\alpha_P$ and $\alpha_D$. If $\alpha_P$ and $\alpha_D$ are too close to zero, then progress slows to a crawl. This can happen if the barrier function does not sufficiently punish points near the boundary of the cone. This issue can be mitigated by increasing $\mu$, which here means choosing $\beta$ closer to $1$. However, then we will descend more slowly along the central path. A judicious choice of $\beta$ at each step is therefore essential for success of the algorithm.

\subsubsection{Corrector step}
For any step length between 0 and 1, the Newton step is guaranteed to decrease the error in the linear equality constraints, but not necessarily in the non-linear equation \eqref{newcomplementarity}. \texttt{SDPB} tries to mitigate this issue by applying a predictor-corrector trick \cite{doi:10.1137/0802028} which works as follows.

One first computes the initial ``predictor" step $(dx_p,dy_p,dX_p,dY_p)$ as discussed above. If one were to simply take this step, the residue $R \colonequals \mu I - X Y$ would become $- dX_p dY_p$, whereas we would like it to be zero. To improve this state of affairs somewhat, we compute another ``corrector" step $(dx_c,dy_c,dX_c,dY_c)$ which is obtained by replacing the residue $R$ in the last equation in \eqref{newtonstep} by
\begin{equation}
\label{Rcorrector}
	R \to \mu_c I - X Y - dX_p dY_p
\end{equation}
and keeping the other equations the same. Then one updates the original variables with the corrector step, using the step length choice discussed above. Notice that in \eqref{Rcorrector} we allowed for a new value $\mu_c$; \texttt{SDPB}'s choice of $\mu_c$ is modeled on that of \texttt{SDPA} \cite{yamashita2003implementation,yamashita2010high,yamashita2012latest} and can be found in detail in \cite{Simmons-Duffin:2015qma}.

We note that it is comparatively cheap to compute the corrector step once the predictor step has been computed. This is because both computations reuse the same matrices $S$ and $Q = B^T S^{-1} B$, whose formation and Cholesky factorization are the most expensive operations in the solver.

\subsubsection{Aside: iterated corrector step}
\label{subsubsec:iteratedcorrector}

\texttt{SDPB}, following \texttt{SDPA}, simply takes one predictor-corrector step for each iteration of the algorithm. We however found that it might be more efficient to repeat the corrector step multiple times, even for standard \texttt{SDPB} runs.

Considering a general setup where we search for the stationary point of a function $L(x)$, if we write the standard Newton step (with unit step length for simplicity) as:
\begin{equation}
	x_{k+1} = x_k - H_{xx}^{-1}(x_k) \cdot \nabla_x L(x_k)
\end{equation}
then the added corrector step would simply be:
\begin{equation}
	x_{k+2} = x_{k+1} - H_{xx}^{-1}(x_k) \cdot \nabla_x L(x_{k+1})\,,
\end{equation}
which importantly ``recycles'' the old Hessian matrix. Repeating the corrector step multiple times (using $H_{xx}^{-1}(x_k)$ for every step) is therefore akin to simple gradient descent, and we can at best expect linear convergence as opposed to the superlinear convergence of proper Newton steps. But computationally speaking it might very well be worthwhile if the computation of $H_{xx}^{-1}$ is expensive, which happens to be the case for our SDPs.

After some experimentation we found that significant speedups could indeed be obtained using an iterative-corrector algorithm for \texttt{SDPB}. We propose a minimal modification of the \texttt{SDPB} algorithm as outlined in algorithm~\ref{alg:iterative-corrector}. Each update $(dx, dy, dX, dY)$ is now computed in two stages: the first stage is the usual predictor step and the second stage is to compute the corrector step iteratively with some general convergence criterion and a limit on the maximum number of steps. 
\begin{algorithm}[h!]
Input $\xi_0 = (x_0, y_0, X_0, Y_0)$\;
Compute $d\xi_p \gets \texttt{PredictorStep}(R)$\;
Initialize $d\xi_c = d\xi_p$\;
\While{$
\left(\text{convergence criterion}\right)
 \,\,\&\,\,
\left(\text{it} \le  \texttt{MAX} \right)$}
{
    Store $d\xi_{c} \gets d\xi_c$\;
    $R_c \gets \mu_c I - X_0 Y_0 - dX_{c} dY_{c} $ \; 
	Update $d\xi_c \gets \texttt{CorrectorStep}(R_c)$ \; 
	it $\gets$ it $+ 1$\;
}
\Return $d\xi_c$\;
\caption{\label{alg:iterative-corrector}The iterative corrector step. 
\texttt{PredictorStep}, \texttt{CorrectorStep}:  standard Interior Point Method subroutines.
}
\end{algorithm}

We have explored two types of convergence criterion. The first strategy uses the residue $R$. We consider a solution to be converging if $R$ continues to decrease with each iteration. The second strategy uses the step length parameters $\min(\alpha_P,\alpha_D)$ as an indicator of convergence. As long as $\min(\alpha_P,\alpha_D)$ is increasing, we consider the solution to be converging. Both the residue $R$ and the step length parameters can function as convergence criteria, but the former has lower computational cost.

In general, we have observed that the iterative corrector steps converge efficiently when the initial point $\xi_0$ is sufficiently close to the true solution for a fixed $\mu$. We have observed linear convergence in actual \texttt{SDPB} runs, in agreement with the above general analysis. It is worth mentioning that there is potential to enhance efficiency by reducing $\mu_c$ during each iterative step.\footnote{We have done some preliminary exploration in this direction. In experiments, it was sometimes possible to decrease the duality gap from $10^{-15}$ to $10^{-30}$ using only corrector iterations (without ever recomputing the Hessian).} We leave the problems of analyzing the convergence rate with varying $\mu$, establishing criteria for how fast to decrease $\mu$, and determining when to stop the corrector iterations for future work.

\subsection{Semidefinite programs with parameters}
\label{sec:withparameters}

As described in the introduction, besides the internal SDP variables
\begin{equation}
	\xi \colonequals (x,y,X,Y),
\end{equation}
the problems of interest depend on additional external parameters $p$ taking values in a parameter space $\cP$. We will assume that the vector of matrices $A_*$ does not depend on $p$, as is adequate for most conformal bootstrap problems, but the other ingredients in the SDP do depend on $p$:
\begin{equation}
	(b, c, B) \quad \to \quad (b(p), c(p), B(p)).
\end{equation}
We furthermore assume that this dependence on $p$ is smooth. Note that this does not imply that the objectives are smooth; see \cite{Reehorst:2021ykw} for an example where the objective is not $C^2$, even though the inputs are $C^\infty$.

We consider two types of optimization problems over $\cP$.  For fixed $p$, let us denote the optimal value of $\xi$ by $\xi_*(p)$. We furthermore define the objective at optimality by
\begin{equation}
	\text{obj}(p) \colonequals c^T(p) x_*(p) = b^T(p) y_*(p).
\end{equation}
The first type of problem is a simple minimization:
\begin{equation}\label{eq:minnvgproblem}
	\text{minimize obj}(p) \text{ over } p \in \cP.
\end{equation}
The second type is a constrained optimization with a linear objective:
\be
	&\text{maximize } v^T p \text{ over } p \in \cP\nn\\
	&\text{such that obj}(p) \leq 0\,.
\label{eq:findboundaryproblem}
\ee

Both types of problem arise naturally in the navigator approach to the conformal bootstrap \cite{Reehorst:2021ykw}. The first type is used to find a (primal) feasible point in $\cP$ or, in bootstrap terminology, a point inside a bootstrap ``island". (For this application, the algorithm can actually terminate as soon as a point with obj$(p) < 0$ is found.) The second type is used to map out the extremal points of bootstrap islands, and consequently provide rigorous upper or lower bounds on the parameters in $\cP$. For this application, it suffices to consider only linear functions $v^T p$ on $\cP$, but the formalism we discuss can straightforwardly be adapted to non-linear objectives. 

Both types of problem fall into the category of \emph{non-linear semidefinite programs}, where one optimizes a function $f(p, x)$ subject to a positive semidefiniteness constraint $X(p,x) \succeq 0$, with $X(\cdot, \cdot)$ a non-affine function. The fundamentals of such programs were analyzed in 1997 \cite{shapiro1997first}. To date several solution methods have been proposed, see for example \cite{leibfritz2002interior,correa2004global}, and we are aware of one publicly available solver \cite{kovcvara2003pennon}. A survey of existing numerical methods can be found in \cite{yamashita2015survey}.

Compared to a general non-linear SDP, our case is special for two reasons. First, our SDPs have additional equality constraints $B^T(p) x = b(p)$, which implies the existence of free variables $y$ on the dual side. Second, our function $X(\cdot,\cdot)$ is non-linear only in the $p$ variables and typically there are significantly fewer of these (about 10-20) than the dimension of $x$ (a few thousand). The numerical experiments reported in this work only pertain to these ``mildly non-linear" semidefinite programs.

\subsubsection{Lagrange function and Newton step}\label{section:Lagrange function and Newton step}
To set up an interior point method for a parameter-dependent SDP, it is natural to start from the same Lagrange function  (\ref{LF}) as before, now viewed as a function of both $p$ and $\xi$:
\begin{equation}
\label{pdependentlagrangian}
	L_\mu(\xi,p) = c^T(p) x + b^T(p) y - x^T B(p) y + \Tr((X - x^T A_*)Y) - \mu \log \det X\,,
\end{equation}
The Newton steps stemming from this function (at finite $\mu$) are easily obtained, but differ slightly for the two types of problem discussed above. The first type just needs stationarity of the Lagrange function:
\begin{equation}\label{eq:minimum}
	\nabla_\xi L_\mu = 0 \text{ and } \nabla_p L_\mu = 0,
\end{equation}
and the step is therefore computed from:
\be
\label{eq:newton-minimize}
  H_{pp} \d p + H_{p\xi} \d \xi &= - \nabla_p L_\mu\nn\\
  H_{\xi p} \d p + H_{\xi\xi} \d \xi &= - \nabla_\xi L_\mu. 
\ee
Here, $H_{pp}, H_{p\xi}, H_{\xi p}, H_{\xi \xi}$ denote schematically the components of the Hessian matrix of second derivatives of the Lagrange function.

An important difference compared to the fixed-$p$ case is that the steps produced by (\ref{eq:newton-minimize}) are no longer guaranteed to step towards a \emph{minimum} of the Lagrangian. Indeed, suppose we have solved the internal component of the problem and therefore $\nabla_\xi L_\mu = 0$. Then we can eliminate $d \xi$ to find:
\begin{equation}
	\delta L_\mu = - \frac{1}{2} \delta p^T \left( H_{pp} - H_{px} H_{xx}^{-1} H_{xp} \right) \delta p + \ldots.
\end{equation}
In this case, we are guaranteed to make progress in the right direction if:
\begin{equation}
\label{eq:reducedhessian}
	H_{pp} - H_{px} H_{xx}^{-1} H_{xp} \succ 0.
\end{equation}
However, this constraint is not automatically satisfied, since we do not assume any particular structure for the hessian and gradient in $\cP$ space. If (\ref{eq:reducedhessian}) is not satisfied, then the na\"ive Newton step might aim for a local saddle point or maximum rather than a minimum. A common method for dealing with this problem is to replace the non-positive-definite (reduced) Hessian (\ref{eq:reducedhessian}) with a positive-definite one. We will outline our procedure for performing this replacement in section \ref{section:Skydiving_Algorithm}.

For the second type of problem, we would like to reach a point with:
\begin{equation}\label{eq:originalboundary}
\nabla_\xi L_\mu = 0 \text{ and } \lambda v = \nabla_p L_\mu \text{ and } L_\mu = 0 \text{ and } \lambda > 0,
\end{equation}
where $\lambda$ is a (scalar) Lagrange multiplier, whose sign is constrained by demanding that we maximize $v^T p$ subject to $L_\mu \leq 0$. In \cite{Reehorst:2021ykw} it was proposed to take the following step:
\be
  \frac12 \left( \d p H_{pp} \d p + \d p H_{p\xi} \d \xi + \d \xi H_{\xi p} \d p + \d \xi H_{\xi\xi} \d \xi \right) + \quad& \nn\\ 
   (\nabla_p L_\mu)^T \d p +  (\nabla_\xi L_\mu)^T \d \xi &= - L_\mu, \label{eq:newton-constrain}\nn\\
    H_{\xi p} \d p + H_{\xi\xi} \d \xi &= - \nabla_\xi L_\mu,\nn\\
    H_{pp} \d p + H_{p\xi} \d \xi &= - \nabla_p L_\mu +\lambda  v.
\ee
The quadratic term on the first line is not present in a regular Newton step. We added it because it is expected to give a more accurate step at negligible computational cost (since it amounts to solving only a single quadratic equation).

It is again useful to consider the case $\nabla_\xi L_\mu = 0$ and eliminate $\d \xi$. The first equation in \eqref{eq:newton-constrain} then becomes:
\begin{equation}
\label{eq:quadraticmodel}
	\frac12 \d p \left(  H_{pp} - H_{p \xi} H_{\xi\xi}^{-1} H_{\xi p} \right) \d p +  (\nabla_p L_\mu)^T \d p = - L_\mu.
\end{equation}
The step will always head towards a zero of this quadratic model. (If there is no zero in sight, \emph{i.e.}, if the solution to the quadratic equation (\ref{eq:quadraticmodel}) in $\de p$ is complex, then we extremize instead by taking its real part.) The other equations can be combined to give:
\begin{equation}
	\lambda \, v^T \delta p = \frac{1}{2} \d p^T \left(H_{pp} - H_{p \xi} H_{\xi\xi}^{-1} H_{\xi p} \right) \delta p - L_\mu.
\end{equation}
Now suppose $L_\mu$ were negative. Then we should make progress towards maximizing the objective, so we would like to have $v^T \delta p > 0$. But the right-hand side of the above equation is only guaranteed to be positive if:
\begin{equation}
	H_{pp} - H_{p \xi} H_{\xi\xi}^{-1} H_{\xi p}  \succ 0\,,
\end{equation}
which is the same constraint we found for the first type of problem. To avoid stepping in an incorrect direction, we should therefore once again modify this matrix to ensure its positivity.

\subsubsection{Newton steps and stalling}
The above equations lead naturally to an interior-point algorithm based on Newton's method. As in the usual algorithm, we first estimate $\mu_\text{est} = \Tr(XY)/\dim(X)$ and set the target $\mu \to \beta \mu_\text{est}$. Then we substitute this value of $\mu$ to compute a Newton-like step $(\d\xi, \d p)$ from equations \eqref{eq:newton-minimize} or \eqref{eq:newton-constrain}. We then update
\begin{equation}
\label{combinednewtonstep}
\xi \leftarrow \xi + \alpha \d\xi, \quad p \leftarrow p + \alpha \d p, 
\end{equation}
where $\alpha = \min(\alpha_P, \alpha_D)$ for $dp$.

In our initial investigations, we implemented essentially this algorithm (although we estimated $H_{pp} - H_{p \xi} H_{\xi\xi}^{-1} H_{\xi p}$ with a BFGS approach). Unfortunately, it did not perform as well as we hoped. The main issue we encountered was that of \emph{stalling}. Concretely we found that, in the course of running the algorithm, $\alpha_P$ and $\alpha_D$ eventually become too small to produce a meaningful step in parameter space.

As we discussed briefly at the end of section~\ref{subsec:review}, small values of $\alpha_P$ and/or $\alpha_D$ indicate that the na\"ive Newton step brings $X$ and/or $Y$ far outside the cone of positive semidefinite matrices. This in turn means that the barrier function $-\mu\log\det(X)$ in (\ref{pdependentlagrangian}) is not sufficiently strong to steer the step towards the interior of the positive cone. It is therefore natural to try to recover from stalling by increasing $\mu$. Although we found that acceptable step sizes can indeed be obtained in that way, doing so is clearly disadvantageous because we would ultimately like to send $\mu \to 0$.

These results highlight a basic tension in this algorithm (and in fact in the use of barrier methods more generally). If one decreases $\mu$ too quickly, one ends up in a stalled situation with infinitesimal step length, and one needs to backtrack by increasing $\mu$. One can avoid hitting the boundary of the cone by decreasing $\mu$ more slowly, but then progress towards the objective might be slow. Although we cannot offer a first-principles derivation of the optimal way to decrease $\mu$ and move in parameter space, in the following we provide a somewhat different algorithm that works well in practice.

Finally let us note that the stalling phenomenon is well-known in other scenarios beyond ours. For example it can occur in standard runs of \texttt{SDPB} where $\beta$ is chosen too close to 0 so that $\mu$ is decreased too quickly. Stalling can also occur in \emph{warm-starting}, where one starts \texttt{SDPB} from a checkpoint or even an optimal solution of an already completed run with similar parameters. In both cases, the problem can be solved in principle by increasing $\mu$, but often at significant computational cost.

\section{Skydiving algorithm}\label{section:Skydiving_Algorithm}

Our main algorithm was developed through trial and error while exploring several representative numerical conformal bootstrap problems. Its principal advantage is that it appears to resolve the stalling problems that plagued na\"ive implementations of the combined Newton step \eqref{combinednewtonstep}.

\begin{figure}[h!]
	\centering
	\includegraphics[width=0.8\textwidth]{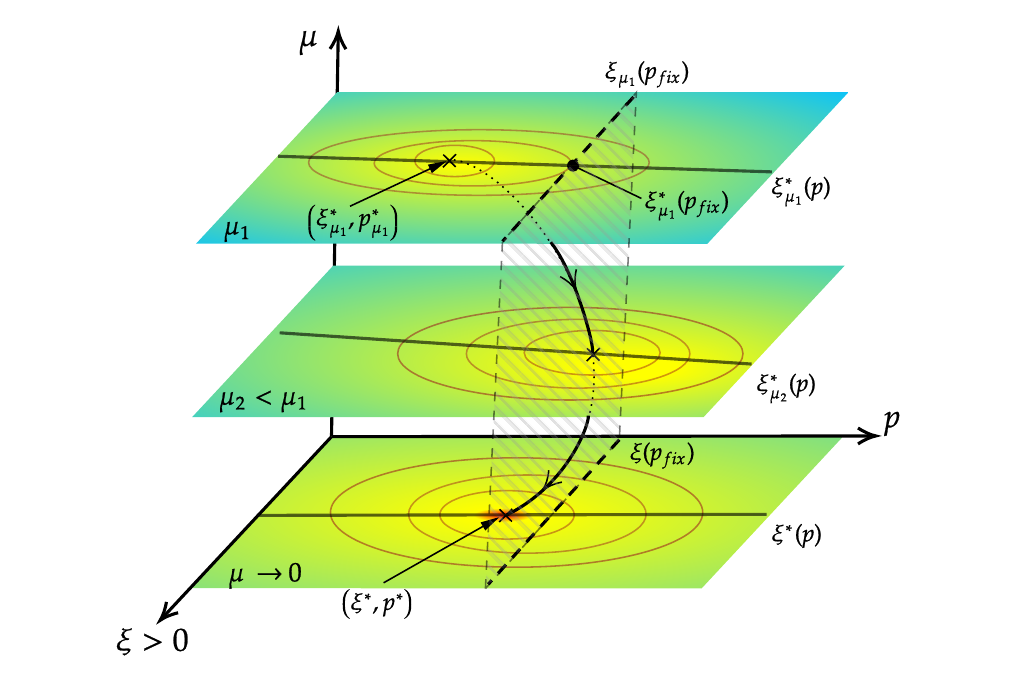}
	\caption[]{\label{fig:visualization}Visualization of central section $\xi_\mu(p)$, optimal solution of standard PDIP run $\xi(p_{fix})$ and global central path $\xi^*_\mu(p^*_\mu)$. A contour map is sketched for the Lagrangian $L_\mu(\xi, p)$ for different values of $\mu$. The usual fixed-$p$ \texttt{SDPB} solves the problem on the shaded vertical  surface, while the skydiving algorithm prioritizes finding the global central path, shown as a curved path traveling through the horizontal surfaces.} 
\end{figure}

Let us first introduce some terminology which we illustrate in figure~\ref{fig:visualization}.

We define the \emph{central section} $\xi^*_\mu(p)$ as the optimal point in $\xi$-space for a given $(p,\mu)$. In figure~\ref{fig:visualization}, the central section is a two-dimensional plane. To avoid clutter, we show it with three solid black  lines indicating the intersection of the central section with the three fixed-$\mu$ surfaces. A physicist might say that $\xi$ is ``on-shell" along the central section, as it satisfies the stationary condition $\nabla_\xi L_\mu = 0$, while $p$ and $\mu$ are fixed. 

Moving along the central section, we can furthermore optimize over $p$, still keeping $\mu$ fixed. Let the optimal value of $p$ at fixed $\mu$ be $p^*_\mu$. The values $(\xi^*_\mu(p^*_\mu),p^*_\mu)$ then trace out a 1-dimensional curve in $(\xi,p)$-space as a function of $\mu$ that we call the \emph{global central path}. The global central path is shown in figure~\ref{fig:visualization} as the black curve interpolating through the fixed $\mu$ surfaces. The optimal point, in both $p$- and $\xi$-space, is reached by sending $\mu \downarrow 0$ along the global central path.

It is also useful to consider a standard primal-dual interior point (PDIP) run in the context of figure~\ref{fig:visualization}. This would correspond to holding $p$ fixed and moving along the shaded straight vertical plane. The intersection of this plane with the central section is conventionally called the ``central path" in the SDP literature. In our context, we call it a ``local central path".

The most important conceptual feature of our algorithm is that it prioritizes proximity to the central section, in the sense that it generically only changes $p$ and $\mu$ if the current value $\xi_\mu(p)$ is sufficiently close to $\xi^*_\mu(p)$. The algorithm therefore incorporates two types of steps: the first are so-called \emph{centering} steps where $p$ and $\mu$ are held fixed and $\xi$ is updated to approach the central section. The second type are full Newton steps, where the aim is to update both $\xi$ and $p$. These Newton steps are more delicate, both because they are sensitive to the non-linearity in $\cP$ and because $\mu$ cannot be decreased too quickly. To avoid stalling, we therefore introduce a new \emph{scanning} technique that we describe in the next subsection. 

As we outlined in the introduction, the original navigator approach of \cite{Reehorst:2021ykw} solves an SDP all the way to optimality, so $\mu \downarrow 0$, before deciding which step to take in $\cP$. With the skydiving approach we instead work at finite $\mu$. The larger we take $\mu$, however, the more our picture of the navigator function over $\cP$ risks being distorted. This implies that our steps in the parameter space risk becoming meaningless if we work at very large $\mu$. For this reason we do not immediately start stepping around in $\cP$. Instead, we initiate our algorithm with a single standard run of the primal-dual interior point (PDIP) algorithm described in the previous section, updating both $\xi$ and $\mu$ until the duality gap is sufficiently small (but not zero). Only then do we enter the main component of the algorithm.

\begin{algorithm}[h!]
initialize $(\xi,p)$\;
\While{not converged}{
\While{duality gap $> \text{dg}_\text{thr}$}{
	update $\xi$ using PDIP step
}
\While{$||\,XY - \mu I \equalscolon R\, ||_{\max} > R_\text{thr}$}{
	update $\xi$ using centering step
}
update $(\xi, p)$ from scanning step
}
\caption{\label{alg:skydiving} Skydiving.}
\end{algorithm}

The broad-strokes algorithm is sketched in algorithm~\ref{alg:skydiving}. In words, while the solution has not converged, we first follow the standard PDIP algorithm to update $\xi$ until $\mu_\text{est} = \Tr(XY)/\text{dim}(X)$ has decreased enough. On the first iteration this typically takes 100-200 PDIP steps, whereas in later iterations the duality gap tends to stay below $\text{dg}_\text{thr}$.
We then enter the second while loop where $\xi$ is updated by a few centering steps until $R$ is small enough. At the end of this loop,  $\xi_\mu(p)$ is brought sufficiently close to $\xi^*_\mu(p)$ so that we declare ourselves to (effectively) be at $\xi^*_\mu(p)$. Finally, we take a scanning step to update both $p$ and $\xi$. 

Of course proximity to the global central path also requires that $\xi$ be both primal and dual feasible, i.e.\ that the linear equality constraints in (\ref{PB}) and (\ref{DB}) are satisfied. In practice, these constraints are usually obeyed once $R \ll 1$. The reason is that often the solver can take a centering step with step length 1 --- i.e.\ a full Newton step. A full Newton step will exactly solve the primal and dual equality constraints, since they are linear equations. Further centering steps will then stay on the primal and dual feasible locus.\footnote{If, for some reason, $R$ happened to be small {\it without} solving the primal dual constraints (for example due to warm-starting with a checkpoint with small $R$), then R will typically become large again in subsequent iterations if they still don't solve the primal dual constraints. Therefore testing $R \ll 1$ usually suffices. A more robust implementation might test the primal and dual constraints as well.}

We now elaborate more on the centering step, deferring details of the scanning step to the next subsection. A centering step is the same as a PDIP step except that $\beta$ is set to $1$ so the target value of $\mu$ is the same as $\mu_\text{est}$. Of course $\mu_\text{est}$ can vary slightly from one centering step to the next, due to the non-linearity of the problem, but we found these changes to be immaterial. In our implementation of centering steps, we  adapt the iterative corrector procedure described in section~\ref{subsubsec:iteratedcorrector}. In particular, we choose the convergence criterion in Algorithm~\ref{alg:iterative-corrector} to be
\be
 \text{convergence criterion} = \left(||R||_{\max} > R_\text{thr}\right) \,\, \& \,\, \left(||R||_{\max} \text{ is decreasing} \right).
\ee
Here, $R_\text{thr}$ is the same parameter as in algorithm \ref{alg:skydiving}. A typical value for the maximum number of corrector iterations \texttt{MAX} is 5. 

\begin{figure}[!ht]
	\centering
	\includegraphics[width=0.8\textwidth]{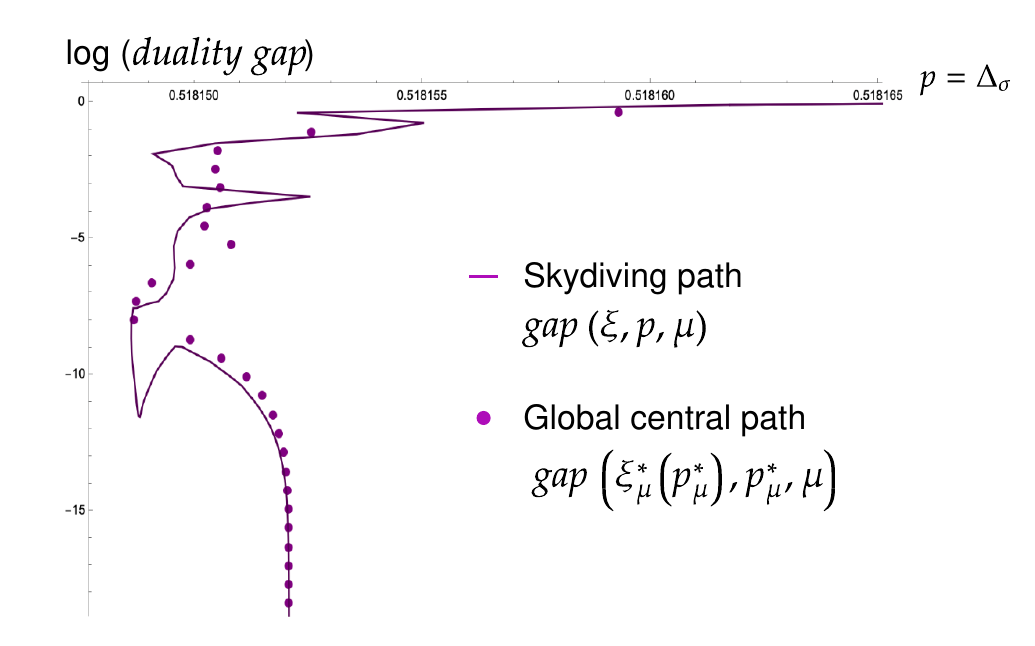}
	\caption[]{\label{fig:centering_eg}An illustration of the proximity between the solver path  $(\xi, p, \mu)$ and the global central path $(\xi^*_\mu(p^*_\mu), p^*_\mu,\mu)$ in the (duality gap)-$p$ plane for a 3d Ising mixed correlator problem. The reader might be surprised at the rather erratic behavior of the global central path when $\log(\text{duality gap}) > -8$. This is due to the unforgiving nature of the navigator function itself for this problem. As we show below in figure \ref{fig:finitemunavigator_smart} it sometimes has two local minima and also contains a nearly flat region.}
\end{figure}

Figure \ref{fig:centering_eg} shows an example skydiving run for the 3d Ising mixed correlator problem described below in section~\ref{section:3dIsingruns}. This is a one-dimensional search over $p = \Delta_\sigma$, with fixed $\Delta_\epsilon$ and $\lambda_{\sigma\sigma\epsilon}/\lambda_{\epsilon\epsilon\epsilon}$. The figure compares the skydiving path (connected lines) with the global central path (individual dots) in the (duality gap)-$p$ plane. (The global central path is obtained by solving equation \eqref{pdependentlagrangian} at fixed $\mu$.)  We see that, despite early fluctuations in the skydiving path, it in general follows the global central path and eventually converges to it. We observed in experiments that without centering steps, the skydiving path often predicts wrong steps in $p$ so that it deviates from the global central path and eventually stalls.

\newpage

\subsection{Scanning}\label{section:scan}
As described above, once a target value of $\mu$ has been set we can solve the linearized equations to obtain a step $(d\xi, dp)$. But is this a good step or are we aiming ``too low" and putting ourselves at risk of stalling? Balancing efficiency and robustness, the scanning algorithm serves to find a good step towards the lowest reasonable target value of $\mu$.  During scanning, we use the step length
\begin{equation}
\alpha \colonequals \min(\alpha_P,\alpha_D)
\end{equation}
as a proxy for whether a step is good or not. The procedure is as follows:
\begin{enumerate}

\item \label{item:asinpdip} As in the PDIP algorithm, the solver begins by choosing a reduction factor $\beta_\text{min}$. We calculate a step $(d\xi, dp)$ and the corresponding $\alpha$ based on the target value
\begin{equation}
\mu = \beta_\text{min} \mu_\text{est}.
\end{equation}
We accept this step if $\alpha$ is big enough, that is if $\alpha > \alpha_\text{thr\_max}$ for some user-defined $\alpha_\text{thr\_max}$.

\item \label{item:ifalphanotbigenough} If $\alpha$ is not big enough, we gradually increase $\beta$ with step size $\Delta \beta$ up to a value $\beta_\text{max}$ (which is typically 1). In other words we use targets:
\begin{equation}
\mu_i = \beta_i \mu_\text{est}, \qquad \beta_\text{min} \leq \beta_1 < \beta_2 < \ldots \leq \beta_\text{max},
\end{equation}
to compute the $i$'th step $(d\xi_i, dp_i)$. Again, we accept a step as soon as $\alpha > \alpha_\text{thr\_max}$. Note that the computation of these steps is cheap because we can recycle the $S$ and $Q$ matrices, just as in the predictor-corrector step.

\item \label{item:ifnotstepisbigenough} If no step is big enough even after reaching $\beta_\text{max}$, we decide to tolerate a smaller but not yet problematic step size. To do so, we store the best step during the search (which does not necessarily occur at $\beta_\text{max}$), and accept it if the corresponding $\alpha_\text{best}$ is bigger than some user-defined parameter $\alpha_\text{thr\_min}$. Clearly, this part of the algorithm is only used if $\alpha_\text{thr\_min} < \alpha_\text{thr\_max}$.

\item \label{item:ifalphaisstilltoosmall} If $\alpha$ is still too small, that is, $\alpha_\text{best} < \alpha_\text{thr\_min} $, then we are at risk of stalling. Based on our experiments, this situation do not occur often, but when it does, we need a recovery mechanism. Our recovery subroutine foregoes the update on $p$ and instead opts for a PDIP step with $\beta = \beta_\text{climbing}>1$ that only updates $\xi$. Another attempt to update $p$ will be made in the next round. In practice, we typically set $\beta_\text{climbing}=2$. 

\end{enumerate}
This scanning procedure is outlined in algorithm~\ref{alg:scanning}.\footnote{If the computed step length $\alpha$ is less than 1 then the step returned by this algorithm would bring us exactly to the point where a primal or dual matrix becomes singular. The issues this may cause are avoided by applying the same step length reduction calculations as in standard \texttt{SDPB} \cite{Simmons-Duffin:2015qma}.}

\begin{algorithm}[h!]
	$\beta \gets \beta_\text{min}$\;
	$\alpha_\text{best} = 0$\;
	\While{$\beta < \beta_\text{max}$}{
		compute $(d\xi, dp)$ from Newton step with target $\mu = \beta \mu_\text{est}$\;
		$\alpha \gets \alpha(d\xi)$\;
		\If{$\alpha > \alpha_\text{thr\_max}$}{
			\Return step $(\alpha \, d\xi, \alpha \, dp)$\;
		}
		\If{$\alpha > \alpha_\text{best}$}{
			$d\xi_\text{best} \gets d\xi$\;
			$\alpha_\text{best} \gets \alpha$\;
		}
		$\beta \gets \beta + \Delta \beta$\;
	}
	\If{$\alpha_\text{best} > \alpha_\text{thr\_min}$}{
		\Return step $(\alpha_\text{best} d\xi_\text{best}, \alpha_\text{best} dp_\text{best})$\;
	}
	compute $d\xi$ using PDIP step with $\beta = \beta_\text{climbing}$\; 
	\Return step $(d\xi,dp = 0)$\;
	\phantom{x\;} 
\caption{\label{alg:scanning}The scanning step to compute $d\xi$ and $dp$.}
\end{algorithm}

Naively, the gradient $\nabla_p L$ and Hessian $H_{pp}$ can be computed with finite differences. However, this is particularly expensive for the Hessian, which in an $n$-dimensional parameter space $\cP$ would require computing $(b(p), c(p), B(p))$ for a constellation of $n(n+1)/2$ points around the actual value of $p$.\footnote{\label{footnote:constellation}By a ``constellation", we mean a collection of SDPs at several nearby points $\{p_0, p_0+\e e_1,\dots, p_0+\e e_n\}$ where $e_i$ is a basis-vector in the $i$-th direction and $\e$ is a small parameter.} In practice, we therefore use the BFGS algorithm discussed below to approximate the Hessian.

Note that the algorithm above currently does not tread ``carefully" in the parameter space $\cP$, and it is possible for it to wander into problematic regions and/or exhibit runaway behavior. One way to remedy runaway behavior would be to include a line search along the computed step $(d\xi, dp)$.
Another possibility is to introduce a bounding box in $\cP$-space. In our experiments, these improvements were not needed, but they might be necessary in future computations.

As with any non-convex optimization, our algorithm is not guaranteed to converge. For example, the algorithm might never find a suitable step and continue to increase $\mu$. In this case, centering steps would still take us toward the central section. However, as we explained before, at large $\mu$ we would not have a reliable picture of the navigator function, and any calculated step could be meaningless. A practical way to address this issue is to introduce an upper limit on the duality gap, beyond which the solver will no longer increase $\mu$.

The PDIP step in algorithm~\ref{alg:scanning} includes a single corrector step, just like the scanning step discussed above. A similar corrector step is implemented whenever $H_{\xi \xi}^{-1}$ was used, for example for $H_{\xi\xi}^{-1} \nabla_p L$. The iterated corrector procedure described in section~\ref{subsubsec:iteratedcorrector} is invoked during centering steps only.  

Finally, let us note that $\alpha_\text{thr\_max}$ should not be set to 1 (typically we set it to 0.6). The reason is that when the algorithm is very close to the global central path (at finite $\mu$), it would simply compute that the step $(d\xi, dp) \approx 0$ if it chooses $\beta=1$ and then take an infinitesimal step with step size $\alpha = 1$. Of course there are many ways to avoid such trivial stalling behavior, but choosing $\alpha_\text{thr\_max}$ strictly below 1 works well in practice. By choosing $\alpha_\text{thr\_max}$ closer to 1, we encourage the algorithm to stay closer to the global central path.

\newpage

\subsection{Modified BFGS algorithm}
For the Newton step in the scanning iterations, we need various gradients and Hessians with respect to $p$. This requires us to compute the $p$-dependence of the input data $b$, $c$, and $B$, as exemplified by the computation of $\grad_p L$, which we recall is given by:
\begin{equation}
\label{gradpL}
	dp^T  \grad_p L = dc^T x + db^T y - x^T dB y\,.
\end{equation}
We currently estimate $d B$, $d b$, and $d c$ using finite differences, which is costly but acceptable. However, to compute $H_{pp}$
\begin{equation}
 	dp^T H_{pp} dp = d^2 c^T x + d^2 b^T y - x^T d^2 B y\,,
 \end{equation}
the cost becomes even less favorable as it scales quadratically with $n$, the dimension of $\cP$. We therefore use a BFGS-type approximation to avoid direct computation of $H_{pp}$. In this subsection, we explain the details.

\subsubsection{Review of standard BFGS}
The core idea behind the Broyden-Fletcher-Goldfarb-Shanno (BFGS) algorithm is to estimate the Hessian matrix using information about the gradient. To briefly review the procedure, let us consider the optimization of a non-linear function $f(x)$. A (quasi-)Newton method gives a sequence of points $x_k$, and at each point we have a gradient $\nabla f(x_k)$. Let us construct the differences
\begin{align}
	s_k &= x_{k} - x_{k-1}\,,\nn\\
	y_k &= \nabla f(x_{k}) - \nabla f (x_{k-1})\,.
\end{align}
The pair $(s,y)$ is insufficient to determine the hessian matrix $H$. Instead, we have a vectorial equation that constrains $H$:
\begin{equation}
	H s = y\,.
\end{equation}
This is called the \emph{secant} equation; it would be exact for infinitesimal step size. BFGS now opts to update a previous estimate $H_{k-1}$ by a linear combination of $y_k y_k^T$ and $s_k s_k^T$, with coefficients determined such that the secant equation holds. This produces
\begin{equation}\label{eq:bfgs_update}
H_{k+1} = H_{k} + \frac{y_k y_k^T}{y_k^T s_k} - \frac{H_k s_k s_k^T H_k^T}{s_k^T H^k s_k }\,.
\end{equation}
With this estimate the new step is $s_{k+1} = - H_{k}^{-1}  \nabla f(x_{k}) $. (In practice, one can directly calculate an update for the \emph{inverse} Hessian, see for example \cite{enwiki:1150228229}, but for our purposes this difference is unimportant.)

\subsubsection{A BFGS-type update for the Hessian}
\label{sec:bfgsforhessian}

The main difference between our setup and the standard BFGS setup is that we need to estimate only one block of the Hessian matrix. Consider two points $(\xi_1, p_1)$ and $(\xi_2, p_2)$ and suppose we know $\grad_p L(\xi_1, p_1)$ and $\grad_p L(\xi_2, p_2)$, the full secant equation reads:
\begin{equation}
\label{fullsecant}
	\begin{pmatrix}
	H_{\xi \xi} & H_{\xi p}\\
	H_{p \xi } & H_{pp}
	\end{pmatrix}
	\begin{pmatrix}
	p_2 - p_1\\
	\xi_2 - \xi_1
	\end{pmatrix}
	=
	\begin{pmatrix}
	\grad_p L(\xi_2, p_2) - \grad_p L(p_1, \xi_1)\\
	\grad_\xi L(\xi_2, p_2) - \grad_\xi L(p_1, \xi_1)
	\end{pmatrix}\,.
\end{equation}
Given that we can easily compute $H_{\xi \xi}$, as well as $H_{\xi p}$ and $H_{p \xi}$ using \eqref{gradpL}, only $H_{pp}$ needs to be estimated, and there are two natural ways of doing so.

\paragraph{Estimating $H_{pp}$ directly}In the first method, we write a separate secant equation for $H_{pp}$ and use it to calculate a BFGS update. Such a secant equation must be of the form:
\begin{equation}
	H_{pp} \cdot (p_2 - p_1) = \grad_p L(\xi_{A}, p_2) - \grad_p L( \xi_{A}, p_1)\,,
\end{equation}
where $A$ can be either 1 or 2, but it is essential that $\xi_{A}$ is the \emph{same} for both gradients. Indeed, in the limit where $(\xi_2, p_2)$ and $(\xi_1, p_1)$ are infinitesimally close, say with a distance $\epsilon$, picking $\xi_2$ for the first gradient and $\xi_1$ for the second gradient leads to an error of order $\epsilon$ on the right-hand side, and therefore an error of order $1$ for the estimate of $H_{pp}$. In contrast, the distinction between $A$ being either $1$ or $2$ is one order higher in $\epsilon$. So to implement this method we need to compute either $\grad_p L(p_2,\xi_1)$ or $\grad_p L(p_1,\xi_2)$, which can be done using the gradient formula \eqref{gradpL}.

\paragraph{Eliminating components} A second method is to use our exact knowledge of $H_{\xi p}$ and $H_{\xi \xi}$ to eliminate the irrelevant components from equation \eqref{fullsecant}. We then get:
\begin{align}
\label{bfgsmethod2}
 	N_{pp} \cdot (p_2 - p_1) = - \grad_p L(\xi_2, p_2) + \grad_p L(\xi_1, p_1) + H_{p \xi} H_{\xi\xi}^{-1} \left( \grad_\xi L(\xi_2, p_2) - \grad_\xi L(\xi_1, p_1) \right)\,,
 \end{align} 
where we introduced:
\be
	N_{pp} &= H_{pp} - H_{p\xi} H_{\xi \xi}^{-1} H_{\xi p}\,.
\ee
and on the right-hand side of equation \eqref{bfgsmethod2} we are omitting the dependence on $(\xi,p)$ of the Hessian components because these give rise to subleading effects.

We can now view equation \eqref{bfgsmethod2} as a separate secant equation, and use a BFGS update for $N_{pp}$ directly from this equation. This is \emph{almost} consistent, but one wrinkle needs to be ironed out: over the course of the algorithm we are actually minimizing slightly \emph{different} Lagrangians because $\mu$ is decreasing! Re-instating this $\mu$-dependence produces:
\begin{equation}
 	N_{pp} \cdot (p_2 - p_1) = - \grad_p L(\xi_2, p_2) + \grad_p L(\xi_1, p_1) + H_{p \xi} H_{\xi\xi}^{-1} \left( \grad_\xi L(p_2,\xi_2,\mu_A) - \grad_\xi L(p_1,\xi_1,\mu_A) \right)\,,	
\end{equation}
where we ignored the $\mu$-dependence in $H_{\xi \xi}$ which again gives rise to higher-order effects. (We recall that $\grad_p L$ does not explicitly depend on $\mu$, which is also why the previous method did not have this wrinkle.) We now need to choose a reference value $\mu_A$. To find an efficient value it is worth looking at the specifics of our algorithm to see which gradients are known. Let us take $\xi_1, \xi_2$ to be the values obtained \emph{after} the centering iterations. Supposing that these iterations succeeded perfectly, we can write:
\begin{align}
\label{gradximuest}
	\grad_\xi L(\xi_1,p_1, \mu_1^\text{est}) &= 0\,,\\
	\grad_\xi L(\xi_2,p_2, \mu_2^\text{est}) &= 0\,,
\end{align}
where we recall that $\mu^\text{est}= \Tr(XY)/\text{dim}(X)$ is the estimated value of $\mu$ at a point $\xi$. Furthermore, in the course of the algorithm we had to compute all the gradients that appear on the right-hand side of the Newton step. Among these we have $\grad_p L(\xi_1, p_1)$ and $\grad_p L(\xi_2, p_2)$, but also  $\grad_\xi L(\xi_1,p_1,\mu_2)$ for some target value $\mu_2$ determined in the scanning step. The Lagrangian is however linear in $\mu$, so using \eqref{gradximuest} we can immediately find one more gradient:
\begin{equation}
	\grad_\xi L(\xi_1, p_1, \mu_2^\text{est}) = \frac{\mu_1 - \mu_2^\text{est}}{\mu_1 - \mu_2} \grad_\xi L(\xi_1, p_1, \mu_2)\,.
\end{equation}
The fraction on the right-hand side, in turn, is roughly equal to the step length $\alpha_1$:
\begin{equation}
	\frac{\mu_1 - \mu_2^\text{est}}{\mu_1 - \mu_2} \approx \alpha_1\,,
\end{equation}
since both sides indicate (at a linearized level) how much progress was made towards the  target $\mu_2$ starting from $(\xi_1, p_1)$. It therefore makes the most sense to evaluate the secant equation for $\mu_2^\text{est}$, where we obtain:
\begin{equation}
	N_{pp} \cdot (p_2 - p_1) \approx - \grad_p L(\xi_2, p_2) + \grad_p L(\xi_1, p_1) - \alpha H_{p \xi} H_{\xi\xi}^{-1} \grad_\xi L(p_1,\xi_1,\mu_2)
\end{equation}
and all the gradients on the right-hand side are known. The BFGS update for $N_{pp}$ based on this secant equation is the algorithm currently implemented in our code.

We stress that this update procedure is technically only applicable if the centering steps are \emph{complete}, in the sense that any residuals in \eqref{gradximuest} translate directly into a mismatch of the update. Our core algorithm, however, does not require this: the scanning step is OK even if the centering steps are reduced or completely omitted.

\subsubsection{Ensuring positivity of the Hessian}

As discussed in section~\ref{section:Lagrange function and Newton step}, positivity of $N_{pp}=H_{pp} - H_{p\xi} H_{\xi \xi}^{-1} H_{\xi p}$ would guarantee progress toward optimality.\footnote{This also provides an argument in favor of using the second method for estimating the Hessian of the previous subsubsection. Namely, if we were to augment our scanning steps with a line search then the we can use the standard Wolfe termination conditions. At the moment this line search is however not yet implemented.} However, positivity is not a-priori guaranteed, and we need a strategy to ensure it. In our implementation, we choose to update our approximation to $N_{pp}$ (using the BFGS-type update discussed above) only when the new $N_{pp}$ would be positive. If positivity would be violated, we continue using the old (non-updated) Hessian.
 
The user can initialize the Hessian either by specifying it explicitly or using the default value
\begin{equation}\label{equ:Hmixed}
N_{pp}^\textrm{default}  = - H_{p \xi} H_{\xi\xi}^{-1} H_{\xi p}\,. 
\end{equation}
This default value generally has the advantage of being similar in scale to the exact $N_{pp}$. 

\subsection{Shifting the Lagrange function}\label{section:Shifted_Lagrangian}

We need a further modification of our algorithm for the second type of optimization (\ref{eq:findboundaryproblem}), which searches for an extremal point of the feasible region in parameter space $\cP$. In such a search, the goal is to find a point where the navigator function \emph{vanishes}:
\begin{equation}
\label{zeromutarget}
	\lim_{\mu \downarrow 0} L_\mu = 0\,.
\end{equation}
A natural question is: how should we interpret this equation at finite $\mu$? In other words, when we work at $\mu > 0$, what is our best guess for the zero locus given by equation \eqref{zeromutarget}? For simplicity, in \eqref{eq:originalboundary} and \eqref{eq:newton-constrain}, we used the simplest finite-$\mu$ version of this equation, namely $L_\mu=0$. However, this is not necessarily the best choice.

For the following discussion it is useful to suppose that we are ``on-shell" in the $\xi$ variables, so we evaluate the Lagrangian at $\xi_\mu^*(p)$. Let us introduce the corresponding on-shell navigator function: 
\begin{equation}
\label{Npdefn}
	N_\mu(p) \colonequals L_\mu(\xi_\mu^*(p),p) = c^T x_\mu^* - \mu \log \det X_\mu^*\,,
\end{equation}
where the second equality holds because the terms multiplying $y$ and $Y$ in equation \eqref{LF} vanish on-shell. In this language, we are interested in the locus $N_{0}(p) = 0$. As explained for example in \cite{Reehorst:2021ykw}, one generically finds that $\lim_{\mu \downarrow 0} \mu \log \det X_\mu^* = 0$ as well (even though $X_\mu^*$ becomes singular) so $N_{0}(p) = c^T x_{0}^*$ which also equals $b^T y_{0}^*$ because the duality gap vanishes.

\begin{figure}[!ht]
\begin{center}
	\includegraphics[width=12cm]{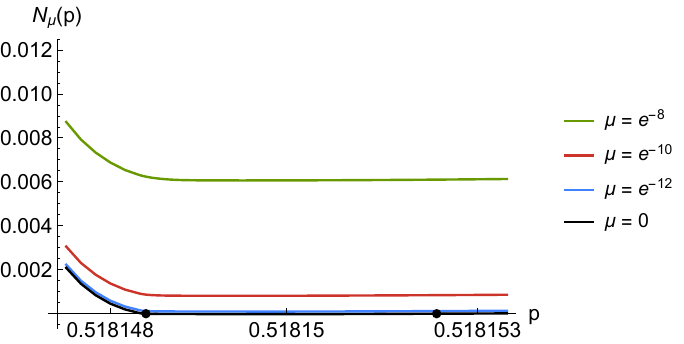}
	\caption{The function $N_\mu(p)$ in a representative example for various values of $\mu$. For $\mu = 0$ there exist two zeroes at the indicated points, but for the other values of $\mu$ shown here the functions are everywhere positive.\label{fig:finitemunavigator_naive}}
\end{center}
\end{figure}

\begin{figure}[!ht]
\begin{center}
	\includegraphics[width=12cm]{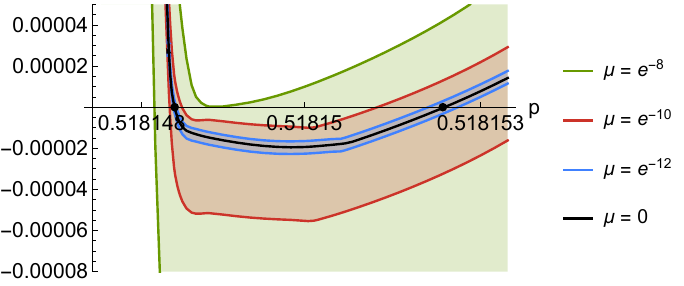}
	\caption{We shaded the region allowed by the bound of equation \eqref{objectivebound} for various values of $\mu$. The black line agrees with the function $N_0(p)$ shown in figure~\ref{fig:finitemunavigator_naive} and, as it should be, lies in between the bounds for each value of $\mu$. Notice that the vertical axis is zoomed in a factor 250 compared to figure~\ref{fig:finitemunavigator_naive}.\label{fig:finitemunavigator_smart}}
\end{center}
\end{figure}

In figure~\ref{fig:finitemunavigator_naive} we show $N_\mu(p)$ in a one-dimensional example for a few different values of $\mu$. (The problem here is that of the three-dimensional Ising model, which is representative of more general conformal bootstrap problems. The parameter $p$ is $\Delta_\sigma$.\footnote{Specifically, the problem is same as the problem in section~\ref{section:3dIsingruns} except we fixed $\Delta_\epsilon=1.412625, \text{arctan}(\lambda_{\epsilon\epsilon\epsilon}/\lambda_{\sigma\sigma\epsilon})=0.9692606$ and used the GFF navigator at $\Lambda=19$. Here we simply use this one parameter example as a demonstration.}) The black line is $N_0(p)$ and, as can be seen in more detail in figure~\ref{fig:finitemunavigator_smart}, there is actually a small negative region, and therefore there exist two points where $N_{0}(p) = 0$. At higher values of $\mu$, however, there are no points with $N_\mu(p) = 0$ at all. Therefore it seems inadvisable to aim for $N_\mu(p) = 0$ as a proxy for $N_0(p) = 0$.

We can improve this state of affairs by working with the primal and dual objectives. Since $\xi_\mu^*(p)$ is optimal at finite $\mu$, it is in particular feasible, which leads immediately to the inequalities:
\begin{equation}
\label{objectivebound}
	b^T y_\mu^* <  b^T y_{0}^*   = c^T x_{0}^* < c^T x_\mu^*, 
\end{equation}
for each $\mu > 0$. We can see this explicitly in figure~\ref{fig:finitemunavigator_smart}. More importantly, we should notice that the vertical axis is zoomed in by a factor 250 compared to figure~\ref{fig:finitemunavigator_naive} so the primal and dual objective at finite $\mu$ provide a much more accurate estimate of $N_0(p)$!

Altogether it therefore makes sense to simply replace the condition $L_\mu = 0$ in equation \eqref{eq:newton-constrain} with an expression of the form:
\begin{equation}
\label{newconstraintarget}
	\widetilde N_\mu(p) \colonequals  (1-\gamma) \, c(p)^T x^*_\mu + \gamma \, b(p)^T y^*_\mu = 0 
\end{equation}
for some parameter $\gamma \in [0,1]$. Solving this equation should bring us much closer to the locus with $N_0(p) = 0$ already at finite $\mu$, and of course for $\mu = 0$ the conditions are equivalent. Using the expression for the duality gap at finite $\mu$
\begin{equation}\label{eq:duality gap mu}
c^T x_\mu^*  -  b^T y_\mu^*   =  \mu \, \text{dim}(X)\,,
\end{equation}
we can rewrite the constraint as, 
\begin{equation}
\label{newconstraintarget_dim}
	\widetilde N_\mu(p) =  c(p)^T x^*_\mu  -  \gamma \,\mu \, \text{dim}(X) = 0\,.
\end{equation}

Generalizing from this on-shell analysis to an off-shell implementation where $\mu = \beta \mu_{\text{est}} $, we find the function
\begin{equation}
	\widetilde L_\mu (\xi, p) \colonequals c(p)^T x  -  \gamma \, \mu_{\text{est}} \, \text{dim}(X) 
\end{equation}
We claim that $\tl L_\mu=0$ provides a better finite-$\mu$ approximation for the boundary of the bootstrap island.
Finally, comparing $\widetilde L_\mu (\xi, p)$ with the original Lagrangian \eqref{pdependentlagrangian}, we have
\begin{equation}
	\widetilde L_\mu (\xi, p) = L_\mu (\xi, p)   -  \gamma \, \mu \, \text{dim}(X)  + \mu  \log \det X + \,\texttt{linear\, errors}.
\end{equation}
Aside from the negligible linear errors from solving for the SDP constraints, $\widetilde L_\mu (\xi, p)$ contains a constant shift $-  \gamma\,\mu \, \text{dim}(X)$, and an $X$-dependent shift $\mu \log \det X$. We emphasize that both shifts are taken purely to locate the zero locus effectively and they do not affect the first two constraints on the gradients in \eqref{eq:originalboundary}. We reiterate the full set of constraints here:\footnote{In the code, this choice is equivalent to specifying the options \texttt{--navigatorWithLogDetX False} and \texttt{--gradientWithLogDetX True}}
\begin{equation}
\label{eq:modifiedboundary}
\nabla_\xi L_\mu = 0 \text{ and } \lambda v = \nabla_p L_\mu \text{ and } \widetilde L_\mu = 0 \text{ and } \lambda > 0. 
\end{equation}

\newpage

\newcommand\Dsig{\Delta_{\sigma}}
\newcommand\Deps{\Delta_{\epsilon}}
\newcommand\D{\Delta}
\newcommand\fT{\f_T}
\newcommand\fsse{\lambda_{\s\s\e}}
\newcommand\feee{\lambda_{\e\e\e}}
\newcommand\fssT{\lambda_{\s\s T}}
\newcommand\feeT{\lambda_{\e\e T}}
\newcommand\calO{\mathcal{O}}

\newcommand\itie{\textit{i.e. }}

\newcommand\zb{\bar{z}}
\def\hb {\bar{h}}

\newcommand\Vp{\vec{V}^{(+)}}
\newcommand\Vm{\vec{V}^{(-)}}
\newcommand\Vth{\vec{V}^{(\theta)}}

\def\cO {{\cal O}}
\def\cOp{\mathcal{O}'}
\def\cN {{\cal N}} 
\def\bbZ {{\mathbb{Z}}} 
\def\dsZ {{\mathbb{Z}}} 

\section{Implementation}\label{sec:Implementation}
\subsection{Software}

We implemented the skydiving algorithm as an open source C++ program called {\tt skydive}. The code is based on \texttt{SDPB} and inherits most of \texttt{SDPB}'s options. The program {\tt skydive} works on a specific SDP and performs the computation in the body of the first {\tt while} loop (the main loop) of algorithm~\ref{alg:skydiving}. The input of {\tt skydive} is the SDP at $p$, a checkpoint $\xi$, a constellation (see footnote~\ref{footnote:constellation}) of SDPs used to compute the gradient, and various options that control the behavior of the computation. The program performs one iteration of the main loop and gives the following output: a step $dp$ in the external parameter space, a checkpoint for a new $\xi$, a gradient of the Lagrangian and a BFGS-updated Hessian.

To set up the full computation, we use external ``driver" programs to drive the main loop of algorithm~\ref{alg:skydiving} and call {\tt skydive}. The external programs set up the bootstrap problem, produce SDPs, call {\tt skydive}, update $p$ and other parameters, and iterate the main loop until optimality is reached.

The source code of our software is available online:
\begin{itemize}
\item The C++ solver \texttt{skydive}:\\ \href{https://github.com/davidsd/sdpb/tree/skydiving\_release}{\tt https://github.com/davidsd/sdpb/tree/skydiving\_release}.
\item The Mathematica framework \texttt{simpleboot}, which can drive the main loop:\\ \href{https://gitlab.com/bootstrapcollaboration/simpleboot}{\tt https://gitlab.com/bootstrapcollaboration/simpleboot}.
\item A Haskell package \texttt{dynamical-sdp} which can also drive the main loop:\\ \href{https://gitlab.com/davidsd/dynamical-sdp}{\tt https://gitlab.com/davidsd/dynamical-sdp}.
\end{itemize}

Two of the present authors (AL and NS) conducted a mini-course on numerical bootstrap methods at the Perimeter Institute in April 2023, which included tutorials on skydiving and associated software tools. The materials from the mini-course can be found online at the following locations
\begin{itemize}
\item Mini course details: \\ \url{https://events.perimeterinstitute.ca/event/45/}.
\item Slides and tutorial code: \\\url{https://gitlab.com/AikeLiu/Bootstrap-Mini-Course}.
\end{itemize}

\subsection{\texttt{skydive} options}\label{section:parameters}

In this section, we describe the input parameters of the \texttt{skydive} program and their relation to the algorithm presented in section~\ref{section:Skydiving_Algorithm}.

\begin{itemize}
\item \texttt{newSdpDirs}: Specifies the path containing a ``constellation" of SDP files that surround the SDP corresponding to the current value of $p$. These files are required to compute the derivatives of the Lagrangian $L(\xi, p)$ (using finite differences), including $\nabla_p L_\mu$, $H_{p\xi}$, and possibly $H_{pp}$ (depending on whether the BFGS algorithm is used). The SDP files should be named \texttt{plus\_i}, \texttt{minus\_i}, and \texttt{sum\_i\_j}, where $ 0 \le i < j < \text{dim}(p) $.

\item \texttt{externalParamInfinitestimal}: The step size $\epsilon$ in $p$-space between each SDP in the constellation and the current center SDP. For example, if the center SDP is computed at $\vec p$, then \texttt{plus\_0} corresponds to the SDP computed at $\vec p + \epsilon \vec{e}_0$, \texttt{minus\_0} corresponds to the SDP computed at $\vec p - \epsilon \vec{e}_0$, and \texttt{sum\_0\_1} corresponds to the SDP computed at $\vec p + \epsilon \vec{e}_0 + \epsilon \vec{e}_1$, where $\vec e_i$ are basis vectors. 

\item \texttt{numExternalParams}: $\text{dim}(p)$, the dimension of the parameter space. 
\item \texttt{totalIterationCount}: The total number of PDIP iterations computed in previous skydiving runs. This option does not affect the actual run, but can be useful for bookkeeping.
\item \texttt{dualityGapUpperLimit}: $dg_\text{thr}$ in algorithm~\ref{alg:skydiving}. If the initial duality gap (either from a checkpoint or from primal/dual initial matrices) is larger than \texttt{dualityGapUpperLimit}, \texttt{skydive} will run PDIP until the duality gap becomes smaller than \texttt{dualityGapUpperLimit}.

\item \texttt{centeringRThreshold}: $R_\text{thr}$ in algorithm~\ref{alg:skydiving}. \texttt{skydive} will run centering steps at a fixed $p$ until $||R||_{\text{max}} \le$ \texttt{centeringRThreshold}. 
\item \texttt{finiteDualityGapTarget}: This option can be used to find the optimal solution at a chosen value of $\mu > 0$, either to minimize the finite $\mu$ navigator function or to extremize $p$ on the finite $\mu$ navigator function.

\item Parameters of $\beta$-scanning, as described in section~\ref{section:scan}:
	\begin{itemize} 
	\item \texttt{betaScanMin}, \texttt{betaScanMax}, \texttt{betaScanStep}: $\beta_{\min}$, $\beta_{\max}$, and $\Delta \beta$ in the scanning routine.
	\item \texttt{stepMaxThreshold}:  $\alpha_\text{thr\_max}$, the threshold to decide whether to accept the step in items~\ref{item:asinpdip} and~\ref{item:ifalphanotbigenough}.
	\item \texttt{stepMinThreshold}:  $\alpha_\text{thr\_min}$, the threshold to decide whether to execute item~\ref{item:ifnotstepisbigenough}.
	\item \texttt{betaClimbing}: $\beta_{\text {climbing}}$, decides which higher $\mu$ the solver should attempt (climb up to) to find $d\xi$ in item~\ref{item:ifalphaisstilltoosmall}.
	\item \texttt{maxClimbingSteps}:  how many times can the solver climb before executing $\xi \rightarrow \xi + d\xi$ in item~\ref{item:ifalphaisstilltoosmall}.
	\end{itemize}
\item Parameters of the modified BFGS algorithm.
	\begin{itemize}
	\item \texttt{useExactHessian}: A boolean value indicating whether to use the full Newton method or the modified BFGS algorithm which does not require the exact values of the Hessian matrix $H_{pp}$. If this value is  \texttt{False}, then the following parameters are required. 
	\item \texttt{prevGradientBFGS}, \texttt{prevExternalStep}, \texttt{prevHessianBFGS}: These three parameters correspond to $ \grad_p L(\xi_1, p_1) - \alpha H_{p \xi} H_{\xi\xi}^{-1} \grad_\xi L(p_1,\xi_1,\mu_2)$, $p_2-p_1$ and $N_{pp}$ in section~\ref{sec:bfgsforhessian}, respectively. If \texttt{prevHessianBFGS} is not specified, \texttt{skydive} will use $N_{pp}^\textrm{default}$ in \eqref{equ:Hmixed} as the default Hessian.
	\end{itemize}
\item Parameters when extremizing  $p$ in a given direction. 
	\begin{itemize}
	\item \texttt{findBoundaryDirection}: If this option is not specified, the program is set to minimize the navigator function in the space of $(\xi, p)$. Otherwise, this option should be a vector $v$ in $p$-space and then the program will maximize $v^T p$. \\
	In practice, if the initial point is far away from the feasible region, it is useful to start from the minimization mode to move to the desired region and turn on  \texttt{findBoundaryDirection} only after dual objective turns negative.
	\item \texttt{findBoundaryObjThreshold}: A threshold parameter that terminates the program when $\mathcal N (p) \le $ \texttt{findBoundaryObjThreshold}. 
	\item \texttt{primalDualObjWeight}: The weight parameter $\gamma \in [0,1]$ of section~\ref{section:Shifted_Lagrangian}. Aside from the possible shift of $\mu  \log \det X$ term (determined by \texttt{navigatorWithLogDetX}), if $\gamma=0$, the navigator function matches with the primal objective, and if $\gamma=1$, the navigator function matches with the dual objective.
	\item \texttt{navigatorWithLogDetX}: A boolean value indicating whether the navigator function is computed with the $\mu  \log \det X$ term. This option determines whether we aim at $ L_\mu=0$ or $ \widetilde L_\mu = 0$ during Newton's steps as described in section~\ref{section:Shifted_Lagrangian}. 
	\item \texttt{gradientWithLogDetX}: A boolean value indicating whether the gradient of the navigator function is computed with the $\mu  \log \det X$ term. This option determines whether we aim at $\nabla_p L_\mu=0$ or $\nabla_p N_\mu = 0$ during Newton's steps as described in section~\ref{section:Shifted_Lagrangian}, where $N_\mu $ is defined defined as \eqref{Npdefn}.
	\end{itemize}
\end{itemize}

\section{Example runs}\label{sec:examples}

In this section, we showcase the performance of the skydiving algorithm applied to the 3d Ising Model and the 3d O(3) Model.\footnote{Some of the runs in this section used older versions of the skydiving algorithm that are slightly different from the one presented in previous sections. However, these differences are immaterial to the results presented here.}

In table \ref{tab:common_parameters}, we specify the values of the \texttt{skydive} parameters that were common throughout the example runs. We provide the values of \texttt{centeringRThreshold} and \texttt{dualityGapUpperLimit} in the subsections below. Some further comments on parameters will be provided in section~\ref{section:tips}.

\begin{table}[!ht]
\centering
\begin{tabular}{|l|c|}
\hline
\texttt{externalParamInfinitestimal} &  $10^{-40}$  \\
\texttt{betaScanMin} & $0.1$  \\
\texttt{betaScanMax} & $1.01$  \\
\texttt{betaScanStep} & $0.1$  \\
\texttt{stepMinThreshold} & $0.1$  \\
\texttt{stepMaxThreshold} & $0.6$  \\
\texttt{maxClimbing} & $1$  \\
\texttt{betaClimbing} & $1.5$  \\
\texttt{primalDualObjWeight} & $0.2$  \\
\texttt{gradientWithLogDetX} & True  \\
\texttt{navigatorWithLogDetX} & False  \\
\hline
\end{tabular}
\caption{{\tt skydive} parameters common for all example runs.}\label{tab:common_parameters}
\end{table}

\subsection{3d Ising island}\label{section:3dIsingruns}

We tested our algorithm on the 3d Ising model mixed bootstrap problem of \cite{Kos:2014bka}, using four-point functions of the lowest-dimension $\mathbb Z_2$-odd operator $\s$ and  $\mathbb Z_2$-even  operator $\epsilon$. The space $\cP$ consists of three parameters: the scaling dimensions $\Dsig$ and $\Deps$, together with the ratio of OPE coefficients $\fsse/\feee$. In the notation of \cite{Kos:2014bka,Reehorst:2021ykw} the navigator setup can be summarized as follows:
\begin{align}
	\textrm{parameters:}&\quad
	p = (\Dsig,\Deps,x=\fsse/\feee), \nonumber\\
	\textrm{SDP objective:}&\quad\mathcal{N}(p) \colonequals \max \vec{\alpha}\cdot\Vp_{\Delta=0,\ell=0}, \nonumber\\
	\textrm{conditions:}
	&\quad\vec{\alpha}\cdot((1,x)\cdot\Vth\cdot(1,x))\ge 0, \nonumber\\
	&\quad\vec{\alpha}\cdot\Vp_{\D,\ell=0}\ge 0 \text{ for } \D \ge 3,\nonumber\\
	&\quad\vec{\alpha}\cdot\Vm_{\D,\ell=0}\ge 0 \text{ for } \D \ge 3,\nonumber\\
	&\quad\vec{\alpha}\cdot\Vp_{\D,\ell}\ge 0 \text{ for } \D \ge \D_\textrm{unitary} \text{ and } \ell=2, 4,\dots,\nonumber\\
	&\quad\vec{\alpha}\cdot\Vm_{\D,\ell}\ge 0 \text{ for } \D \ge \D_\textrm{unitary} \text{ and } \ell=1, 2,\dots,\nonumber\\
	&\quad\vec{\alpha}\cdot M_\Sigma = 1,
	\label{cond:nvg3par}
\end{align}
where
\be\label{eq:Vtheta}
 \Vth=\Vp_{\Deps,0} + \Vm_{\Dsig,0} \otimes \left(\begin{array}{cc}1 &0 \\0 &0\end{array}\right).
\ee
Here $\Vp$ and $\Vm$ are the $\mathbb Z_2$-even and $\mathbb Z_2$-odd crossing vectors introduced in \cite{Kos:2014bka} and $M_\Sigma$ imposes a normalization corresponding to the $\Sigma$-navigator described in \cite{Reehorst:2021ykw}.\footnote{The specific choice of the vector does not affect the boundary of the feasible region and has minimal impact on the numerical results.} All these conditions depend non-linearly on the parameters $p$, which is implicit in our notation. A point $p \in \cP$ is excluded if the objective is positive and all the conditions are met. 

With this setup we performed three runs of the skydiving algorithm, with the following goals:
\begin{align}
	\textrm{runs 1 and 3:} &\qquad \max \Delta_\sigma \textrm{ such that } \mathcal N(p) \leq 0 \nonumber\\
	\textrm{run 2:} &\qquad \min \mathcal N(p)
\end{align}
Note that for runs 1 and 3 we aim for the rightmost tip of the Ising island and not the larger ``continent" where $\Delta_\sigma$ is unbounded. The minimization of $\mathcal N(p)$ for run 2 is less physically relevant, but it might be a good predictor for the true values of the Ising CFT data as discussed in \cite{Reehorst:2021ykw}.

The details of each run are presented in table~\ref{tab:Isingruns}. The most important qualitative difference between runs 1 and 3 is that the former was done at bootstrap derivative order $\Lambda = 19$ and the latter at $\Lambda = 35$.\footnote{See \cite{Kos:2015mba} for the definition of the parameter $\Lambda$.} We terminated the runs when the first 12 digits of each component of $p_{\text{final}}$ stabilized.

\begin{table}[!ht]
\begin{center}
\begin{tabular}{|l|l|l|l|}
\hline
run                                         & 1                    & 2                    & 3                    \\ \hline
initial point                                & $p_1$                & $p_2$                & $p_3$                \\ \hline
goal                                         & max $\Delta_\sigma$  & min navigator        & max $\Delta_\sigma$  \\ \hline
precision                                    & 448                  & 448                  & 768                  \\ \hline
$\Lambda$                                    & 19                   & 19                   & 35                   \\ \hline
$\kappa$                                     & 14                   & 14                   & 32                   \\ \hline
spins                                        & $S_{19}$             & $S_{19}$             & $S_{35}$             \\ \hline
Initial Hessian                              & $H_\text{init}$      & none                 & none           \\ \hline
$\texttt{dualityGapUpperLimit}$ for 1st SDP  & $10^{-3}$            & $10^{-6}$            & $10^{-15}$           \\ \hline
$\texttt{dualityGapUpperLimit}$ for the rest & $10^{-3}$            & none                 & none                 \\ \hline
$\texttt{centeringRThreshold}$               & $10^{-10}$           & $10^{-10}$           & $10^{-10}$           \\ \hline\hline
final point                                  & $p_{1,\text{final}}$ & $p_{2,\text{final}}$ & $p_{3,\text{final}}$ \\ \hline
total number of {\tt skydive}  calls: $d\xi, dp$               & 236                  & 170                  & 133                  \\ \hline
total number of PDIP  iterations:   $d\xi$ & 580                  & 411                  & 601                  \\ \hline
total number of Newton steps     & 816                  & 581                  & 734                  \\ \hline
\end{tabular}
\caption{\label{tab:Isingruns}Individual parameter setup for example runs on 3d Ising model mixed bootstrap problem and the corresponding results. See the main text for the explanation of several the entries.
}
\end{center}
\end{table}

The symbols in the table correspond to the following values. First, the initial points for each run were
\begin{align}
p_1 &=(0.515, 1.4, 0.5)\nonumber\\
p_2 &=(0.51814, 1.4121, 0.686)\nonumber\\
p_3 &=(0.5181496477062039, 1.4126328939948094, 0.6863837285229739)\nonumber\,,
\end{align}
for reasons explained below, the initial Hessian for the first run was 
\begin{align}\label{equ:Ising_Hinit}
H_\text{init} &=\text{diag}(111.7955564168356,1.184293042423152,0.8299842690871352)\,,
\end{align}
and the spin sets $S_\Lambda$ correspond to
\begin{align}
S_{19}&=\{0,..., 26\}\cup\{49,50\}\nonumber\\
S_{35}&=\{0,..., 44\}\cup\{47, 48, 51, 52, 55, 56, 59, 60, 63, 64, 67, 68, 71, 72, 75, 76, 79, 80\}\,.
\end{align} 
As for the results, in the first run we found that $\Delta_\sigma$ is maximized at 
\begin{align}
p_{1,\text{final}}& =(0.518193035759,1.41299963388,0.686124688310)\,,\nonumber
\end{align}
in the second run that the navigator function is minimized at 
\begin{align}
p_{2,\text{final}}&=(0.518134931547,1.41244135167,0.686615119182)\,,\nonumber
\end{align}
and in the third run with $\Lambda = 35$ that $\Delta_\sigma$ is maximized at  
 \begin{align}
 p_{3,\text{final}}&=(0.518151700083,1.41264851228,0.686376348482)\,.
 \end{align}

\begin{figure}[!ht]
\centering
\begin{subfigure}[b]{0.45\textwidth}
     \includegraphics[width=1.1\textwidth]{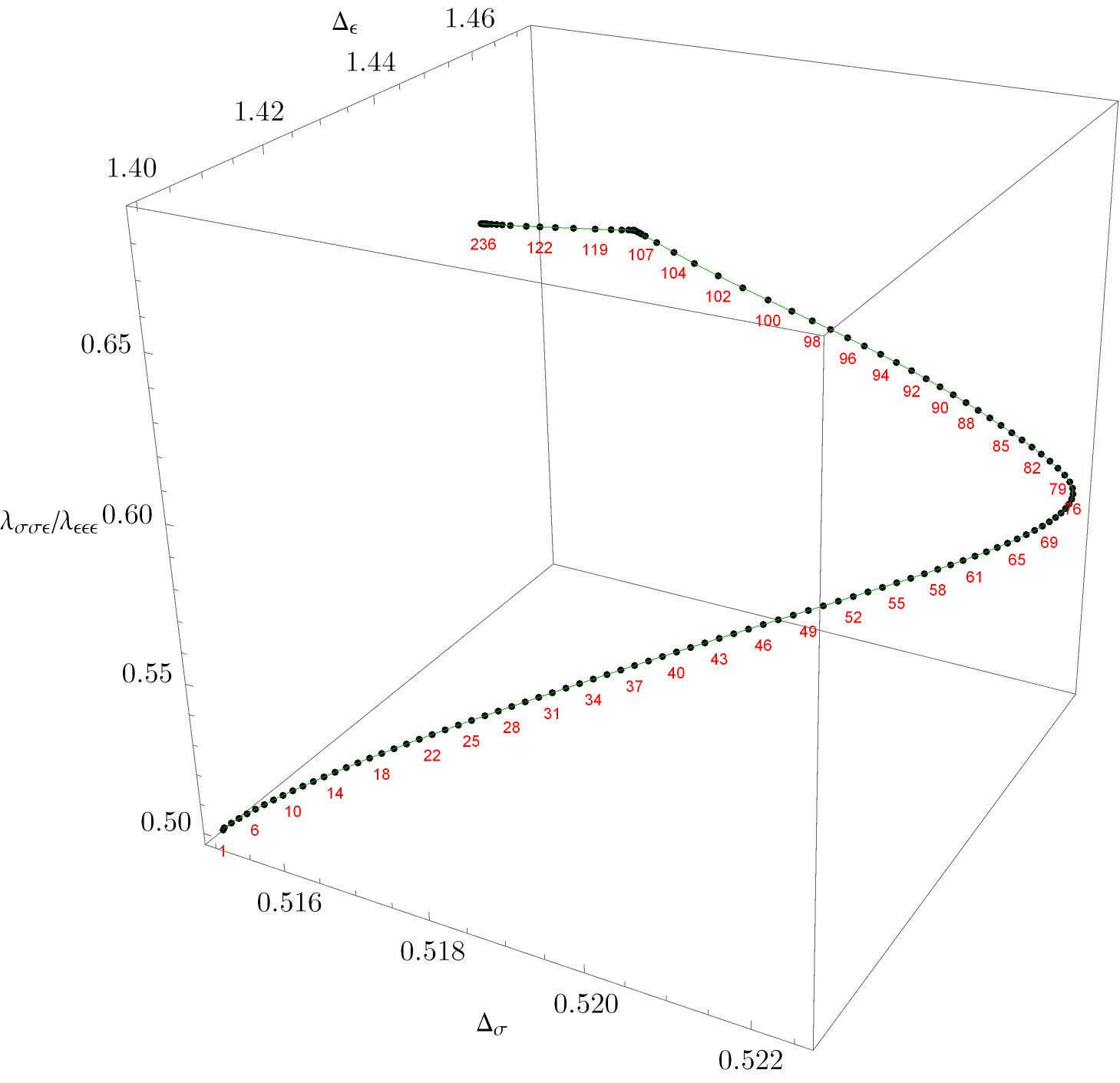}
    \caption{\label{fig:Ising3param_path_3d}}
\end{subfigure}\quad\quad
\begin{subfigure}[b]{0.4\textwidth}
     \includegraphics[width=1.2\textwidth]{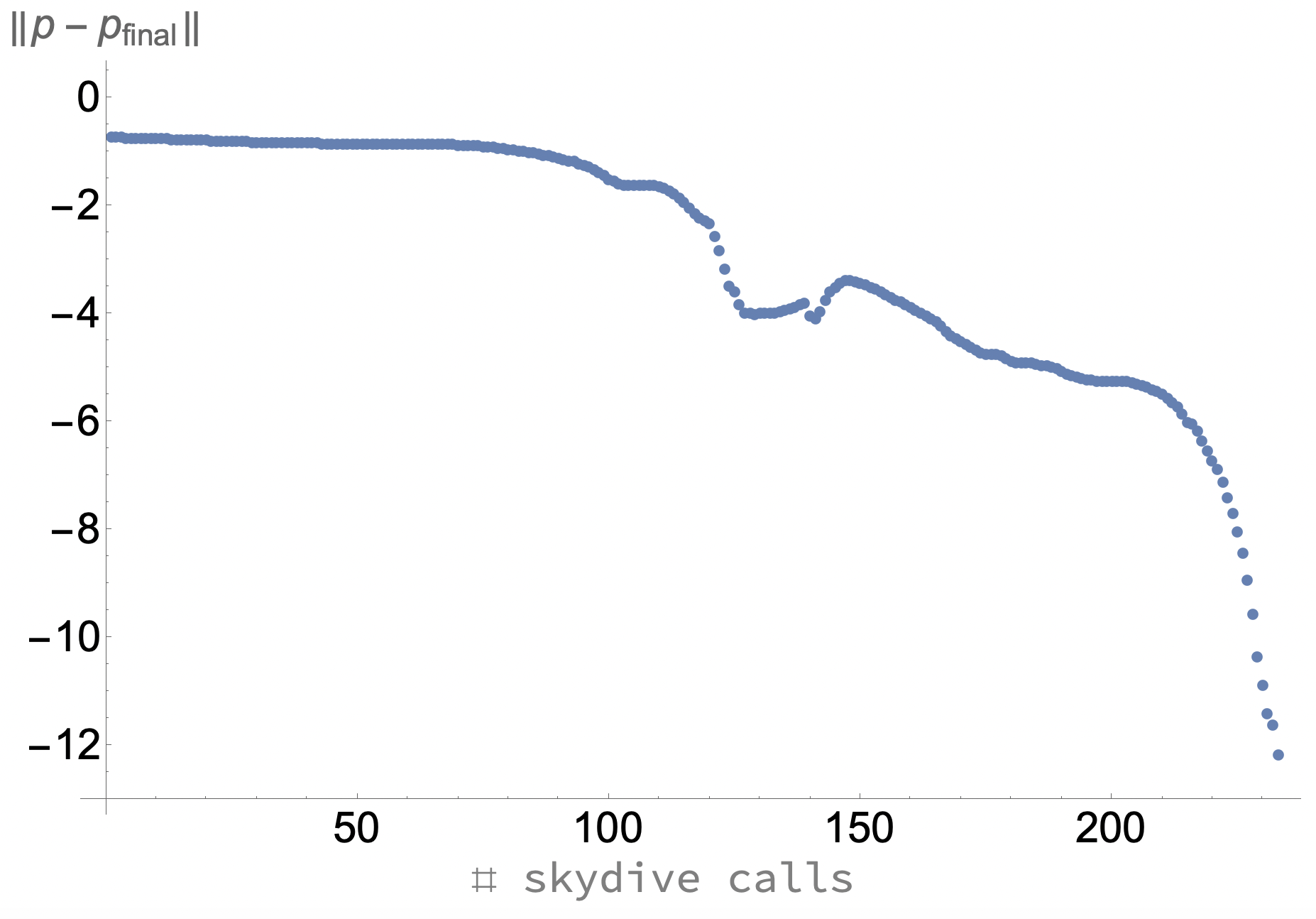}
    \caption{\label{fig:Ising3param_distance_vs_SDP}}
\end{subfigure}
\caption{\label{fig:Ising3param_path}The path of a skydiving run 1 in table \ref{tab:Isingruns}. On the left, the path in the three-dimensional parameter space. On the right, the distance to the final point as a function of the number of \texttt{skydive} steps.}
\end{figure}

Let us offer some further comments on the setup for run 1. Here the initial point $p_1$ was deliberately chosen to lie rather far from the physically interesting region. In fact, with the standard navigator setup of \cite{Reehorst:2021ykw} we were unable to make any run starting at $p_1$ converge to the Ising island. With our new algorithm, however, this turned out to be possible (and subsequently we could maximize $\Delta_\sigma$). It did unfortunately still require some trial and error. For example, we chose the initial Hessian matrix to be as in equation \eqref{equ:Ising_Hinit} --- but we expect that similar initial values will also work. We also set \texttt{dualityGapUpperLimit} to $10^{-3}$ throughout the run in order to prevent it from increasing $\mu$ and diverting to the continent in the initial stages. Altogether these tweaks might be useful in the future, where we would like to better control the behavior of the algorithm as it steps around in $\cP$.

\begin{figure}[!ht]
	\centering
	\includegraphics[width=0.6\textwidth]{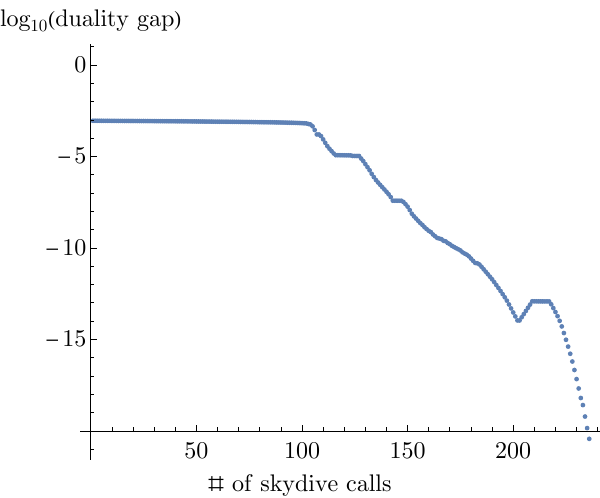}
	\caption[]{\label{fig:Ising3param_dGap_vs_SDP}Duality gap in each {\tt skydive} call during run 1 in table~\ref{tab:Isingruns} which maximizes $\Delta_\sigma$ from point $p_1$ at $\Lambda = 19$.}
\end{figure}

\begin{figure}[!ht]
	\centering
	\includegraphics[width=0.9\textwidth]{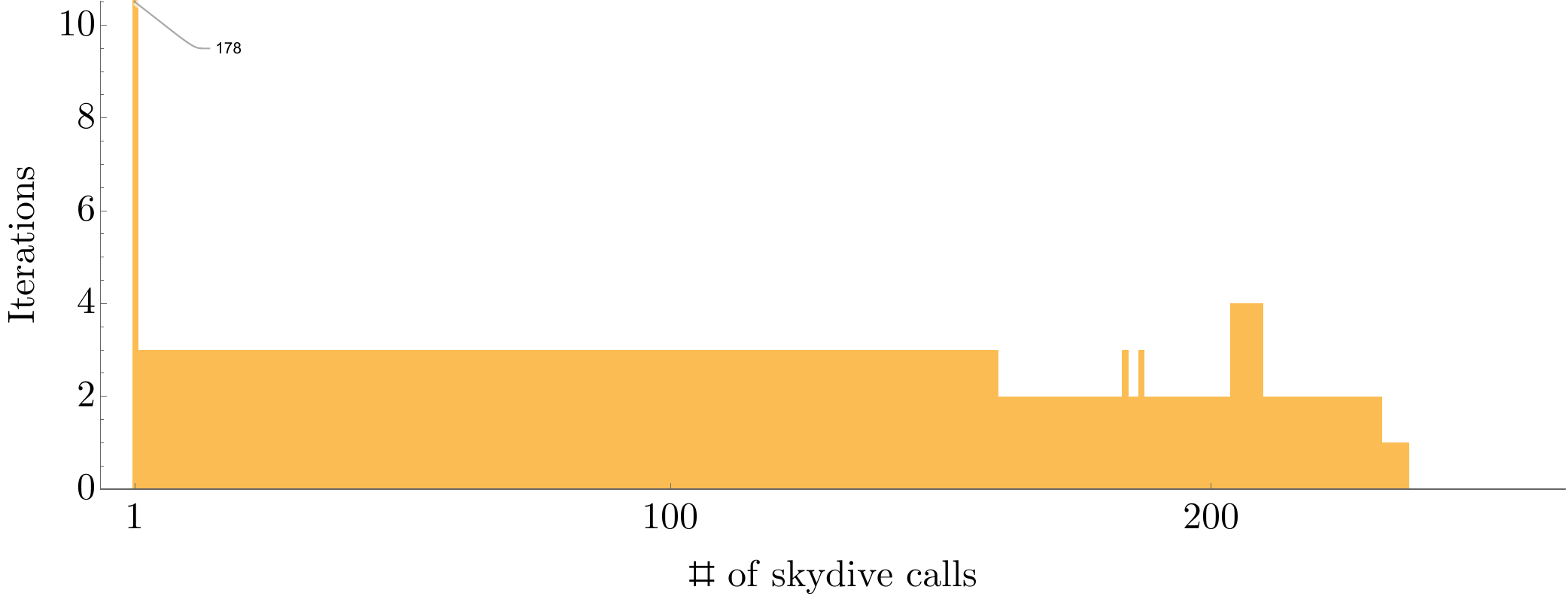}
	\caption[]{\label{fig:ising_iters_vs_Npoint}The number of Newton iterations spent in each {\tt skydive call} for the skydiving run of $p_1$. For each call one Newton iteration is necessary for the scanning step, and the remaining Newton iterations are due to scanning steps or, as in the first \texttt{skydive} call, standard PDIP steps used to lower the duality gap.}
\end{figure}

Let us now discuss the progression of the skydiving algorithm during run 1. As indicated in table \ref{tab:Isingruns}, run 1 took 236 calls to \texttt{skydive} to converge. In figures \ref{fig:Ising3param_path_3d}, \ref{fig:Ising3param_distance_vs_SDP}, \ref{fig:Ising3param_dGap_vs_SDP} and \ref{fig:ising_iters_vs_Npoint} we plot various quantities as a function of the number of steps. For brevity, we do not include plots for the other two runs. They were qualitatively similar, up to one exception that we discuss below.

Let us first discuss convergence. Figure \ref{fig:Ising3param_path_3d} shows the path taken in the three-dimensional parameter space. The distance to the final point is shown in figure \ref{fig:Ising3param_distance_vs_SDP}. Finally, figure \ref{fig:Ising3param_dGap_vs_SDP} shows how the duality gap decreases to zero.\footnote{After we finished computations for those plots, we discovered a small bug in our code: during the scanning step, the code used the value of the navigator function at $\beta=1$ to solve for the boundary, but in fact it should use the extrapolated value at the scanned $\beta$. After fixing this bug, typical runs had about 5\% fewer iterations than the ones presented in this paper.}

For the first 100 iterations or so we observe slow but steady movement towards the Ising island in $\cP$, whereas the duality gap stays essentially flat. During this time, the algorithm would have frequently preferred to take a climbing step (with $\beta > 1$) but it is not allowed to do so because the duality gap has reached our set \texttt{dualityGapUpperLimit} of $10^{-3}$. It therefore settles instead for a full $(d\xi, dp)$ step, even though the corresponding step length $\alpha$ is very small. We stress that this is exactly the desired behavior at this stage, since the bigger steps in $\cP$ do not aid convergence toward the Ising island. It is this initial part that is atypical compared to the other two runs.

After about the 100th iteration, things accelerate. The duality gap begins to decrease and we observe approximate linear convergence. At about the 200th \texttt{skydive} call we observe an \emph{increase} in the duality gap for several steps. This is the stalling recovery mechanism at work, where the algorithm chose $\beta > 1$ to ensure a reasonable step size. In the very last stage, we finally observe (slightly) superlinear convergence. Together, these figures vividly illustrate how the different subroutines described in section~\ref{section:Skydiving_Algorithm} work cooperatively in a complete run of the skydiving algorithm.

The dramatic improvement of the computational cost of our algorithm over a standard navigator function run is visualized in figure \ref{fig:ising_iters_vs_Npoint}. To understand it, we should first note that using \texttt{skydive} does not seem to require a much larger number of steps $dp$ than the standard navigator approach of \cite{Reehorst:2021ykw}. For example, a problem like our run 2 (navigator minimization over 3 parameters at $\Lambda = 19$) would typically have required about 100 steps (\cite{Reehorst:2021ykw}, figure 22) whereas it took us 173 steps (see table \ref{tab:Isingruns}). Therefore, to compare the algorithms it is meaningful to just compare the computational cost of obtaining a single $dp$, i.e., of a single call to \texttt{skydive}.

The costliest operation in our algorithm is the inversion of the Hessian $H_{\xi\xi}$ to produce an update $d\xi$. In figure \ref{fig:ising_iters_vs_Npoint} we plot the number of $d\xi$'s computed for each call to \texttt{skydive} during run 1. In the first iteration, we begin with a standard PDIP run to bring the duality gap below \texttt{dualityGapUpperLimit}, which takes about 200 Hessian inversions. Afterwards, we however only compute $d\xi$ a handful of times for each step $dp$. More precisely, we compute $d\xi$ once for each of the (typically 2 or 3) centering steps, plus once more in conjunction with $dp$ in the scanning step. In the traditional navigator approach, on the other hand, an entire SDP was solved to optimality before calculating $dp$. This would mean that the new algorithm reduces about 200 Hessian inversions to just 2 or 3!

The final rows in table \ref{tab:Isingruns} show that the same speedup was realized for the other two runs. If we take run 2 as an example: the standard navigator approach would need about $2 \cdot 10^4$ Hessian inversions (200 steps $d\xi$ for each of the 100 steps $dp$) whereas we reached the same optimal point with only 581 such operations (3\% as many).

\subsubsection{Computational resources}

In this section we give more details on the computational resources used for run 1. We performed the computation on a computer with 40 CPU cores.\footnote{The computer is a node with Intel(R) Xeon(R) Gold 6148 2.40GHz CPU on the Symmetry cluster of the Perimeter Institute. We thank the Perimeter Institute for the computational resources.} The computation uses \texttt{simpleboot} in high efficiency mode, where all heavy computations are done in C++ programs \cite{Su:2022xnj}. Within one iteration of the main loop of Algorithm \ref{alg:skydiving}, the heavy computations are: (1) computation of convolved conformal blocks (using the \texttt{scalar\_blocks\_mod} program); (2) creation of the files defining SDPs (using the \texttt{sdp2input\_mod} program); (3) running the \texttt{skydive} program. Other than these three C++ programs, the rest costs a negligible amount of time. 

During each iteration, there are 12 \texttt{scalar\_blocks\_mod} calls and 4 \texttt{sdp2input\_mod} calls (corresponding to 4 SDPs). The precision of both programs was set to 448 binary digits. To efficiently use the 40 cores, we launched batches of 6 calls of \texttt{scalar\_blocks\_mod} simultaneously and each call had 7 OpenMP threads. After 12 \texttt{scalar\_blocks\_mod} calls were finished, we launched 4 \texttt{sdp2input\_mod} calls simultaneously and each call had 10 MPI ranks. We found this is the optimal scheme to produce the SDPs. Finally, \texttt{skydive} was launched with 40 MPI ranks. The total computation from $p_1$ took 13696 seconds, that is, less than four hours, among which 40\% is spent on \texttt{scalar\_blocks\_mod}, 30\% on \texttt{sdp2input\_mod}, 30\% on \texttt{skydive}. 

If we look at the cost of the \texttt{skydive} calls alone, on average 48\% of the time was spent on reading and writing data from and to disk. Furthermore, we found that CPU efficiency of the total computation is 31\% , i.e., 69\% of the time the CPUs were idle. 

We would like to emphasize that, in this run, we have already carefully optimized the SDP-generating part of the run. However, the total time spent running \texttt{skydive} is still less than the time spent on generating the SDPs. Thanks to the skydiving algorithm, solving many different SDPs is no longer the bottleneck of this bootstrap computation!

\subsection{\texorpdfstring{$O(3)$}{O(3)} tiptop search}

In this section, we describe an application of skydiving to the 3d critical O(3) model. A previous bootstrap exploration of this model \cite{Chester:2021} studied correlation functions of three scalar operators $(v, s, t)$ with lowest dimension in the vector, scalar, and traceless symmetric tensor representations of $O(3)$, respectively. Stability of the $O(3)$ model under RG flow depends on the scaling dimension of the lowest-dimension scalar $t_4$ in the rank-4 tensor representation of $O(3)$. If $\Delta_{t_4} <3$, then the $O(3)$ fixed point is unstable.

The authors of \cite{Chester:2021} obtained an upper bound on $\De_{t_4}$ only slightly less than 3, thereby rigorously demonstrating instability of the $O(3)$ model. That computation was however very costly. It involved sampling many points in an 8-dimensional parameter space $\cP$. Furthermore, the proximity of the best upper bound to 3 made the search rather delicate.
It is therefore an ideal example to test whether skydiving can be more efficient.

The bootstrap problem from \cite{Chester:2021} is as follows:
\begin{align}
	\textrm{parameters:}&\quad p = \p{\Delta_v,\Delta_s,\Delta_t,\Delta_{t_4},\frac{ \lambda_{vtv}}{\lambda_{vvs}},\frac{\lambda_{tts}}{\lambda_{vvs}},\frac{\lambda_{ttt}}{\lambda_{vvs}},\frac{\lambda_{sss}}{\lambda_{vvs}}}, \nonumber\\
	\textrm{SDP objective:}&\quad\vec{\alpha}\cdot V^{([0,+])}_{\Delta=0,\ell=0}, \nonumber\\
	\textrm{goal:}&\quad \text{maximize  }  \Delta_{t_4} \text{ while SDP objective} \le 0 \nonumber\\
	\textrm{conditions:}
	&\quad\vec{\alpha}\cdot((1,\lambda_{vtv},\lambda_{tts},\lambda_{ttt},\lambda_{sss})\cdot\Vth\cdot(1,\lambda_{vtv},\lambda_{tts},\lambda_{ttt},\lambda_{sss}))\ge 0, \nonumber\\
	&\quad\vec{\alpha}\cdot V^{[1,-]}_{\D,\ell=0}\ge 0 \text{ for } \D \ge 3,\nonumber\\
	&\quad\vec{\alpha}\cdot V^{[4,+]}_{\D,\ell=0}\ge 0 \text{ for } \D \ge \Delta_{t_4},\nonumber\\
	&\quad\vec{\alpha}\cdot V^{[2,+]}_{\D,\ell=0}\ge 0 \text{ for } \D \ge 3,\nonumber\\
	&\quad\vec{\alpha}\cdot V^{[0,+]}_{\D,\ell=0}\ge 0 \text{ for } \D \ge 3,\nonumber\\
    &\quad\vec{\alpha}\cdot V^{\textrm{EMT}}_{\D=3,\ell=2}\ge 0 ,\nonumber\\
	&\quad\vec{\alpha}\cdot V^{R}\ge 0 \text{ for } \D \ge \D_\textrm{unitary}+\delta \text{ and all other $R$ and spins $\ell$}.
	\label{cond:O3tiptop}
\end{align}
Here $\Vth$ represents the crossing vector that has the external operators $(v, s, t)$ appearing as an intermediate operator, $\delta$ is a small twist gap set to $10^{-7}$, and $V^{R}$ denotes the crossing vector for the representation $R$.
The representations appearing in this setup are 
$$
{[1,-],[4,+],[2,+],[2,-],[3,+],[3,-],[4,+]}\,,
$$
where the notation $[n,\pm]$ refers to the O(3) rank $n$ traceless symmetric tensor with $O(3)$ parity $\pm$. $V^{\textrm{EMT}}$ represents the crossing vector for the stress tensor, using OPE coefficient ratios determined by the Ward identity.\footnote{The specific details of all the crossing vectors may depend on conventions. In practice, we used the \texttt{autoboot} \cite{Go:2019lke} package to generate these crossing vectors.}

The only modification between the setup of \eqref{cond:O3tiptop} and ours is that we switched from feasibility mode to navigator mode. The problem is then exactly analogous to the Ising case of the previous subsection: we define $\mathcal N(p)$ as the maximum of the SDP objective and, after imposing an suitable normalization $\vec \alpha \cdot M_{\Sigma} = 1$, we slightly reformulate the goal as maximizing $\Delta_{t_4}$ while $\mathcal N(p) \leq 0$.

We conducted three \texttt{skydive} runs at $\Lambda=19$, each starting from a different initial point. The parameters and performance statistics are summarized in table~\ref{tab:O3tiptopruns}. Again, the symbols in the table correspond to the following values. First, the initial points for each run were
\begin{align}
p_1&=(0.518957,1.59539,1.2097,2.97757, 3.0461,2.4233,3.98997,0.557463)\nonumber\\
p_2&=(0.52,1.6,1.21,3,3,2.5,4,0.5)\nonumber\\
p_3&=(0.517,1.6,1.21,3,3,2.5,4,0.5)\nonumber
\end{align}
 All three runs reached the optimum point:
\begin{align}
p_\text{final}&=(0.519124, 1.59715, 1.21049, 2.99985, 3.04764, 2.42518, 3.99318, 0.560976), 
\end{align}
from which we in particular conclude that
\begin{equation}
	 \Delta_{t_4} \lesssim 2.99985\,, 
\end{equation}
and therefore the $O(3)$ model is unstable.\footnote{Note that this upper bound is weaker than the one in \cite{Chester:2021}. This is because we have used fewer derivatives $(\Lambda=19)$ as opposed to $\Lambda=35$ in \cite{Chester:2021}.}

\begin{table}[!ht]
\begin{center}
\begin{tabular}{|l|l|l|l|}
\hline
run & 1                                                                                                         & 2                                                                      & 3                                                                       \\ \hline
initial point    & $p_1$ & $p_2$ & $p_3$ \\ \hline
\texttt{dualityGapUpperLimit} for 1st SDP         & 0.0001                                                                                                    & 0.001                                                                  & 0.001                                                                   \\ \hline
\texttt{dualityGapUpperLimit} for the rest      & none                                                                                                      & 0.001                                                                  & 0.001                                                                   \\ \hline
\texttt{centeringRThreshold}         & $10^{-10}$                                                                                                & $10^{-40}$                                                             & $10^{-10}$\\
\hline
\hline
total number of \texttt{skydive} calls: $d\xi, dp$ & 254   &401 & 515 \\ \hline
total number of PDIP iterations: $d\xi$ & 577 & 1446 & 1058\\ \hline
total number of Newton steps & 831& 1847& 1573\\ \hline
\end{tabular}
\caption{\label{tab:O3tiptopruns}Skydiving computations of the O(3) tiptop search starting from different initial points at $\Lambda = 19$. The initial points $p_i$ are given in the main text.}
\end{center}
\end{table}
\begin{figure}[!ht]
	\centering
	\includegraphics[width=.5\textwidth]{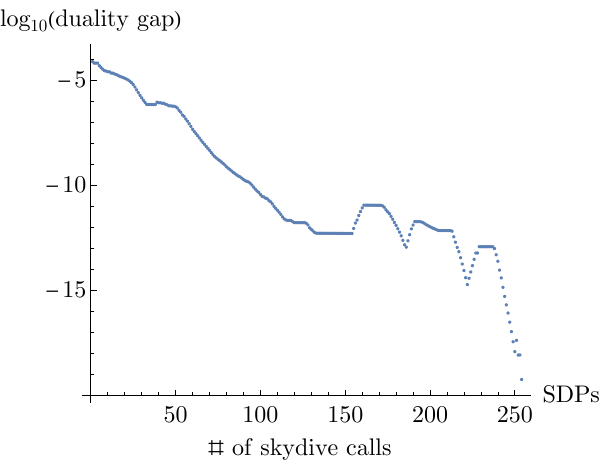}
	\caption[]{\label{fig:LogMu_vs_Npoint}Duality gap in each {\tt skydive} call during run 1 in table~\ref{tab:O3tiptopruns} which maximizes $\Delta_{t_4}$ from point $p_1$ at $\Lambda = 19$.}

\end{figure}

Again, let us provide more details on run 1 as an example. 

 In figure~\ref{fig:LogMu_vs_Npoint}, we illustrate how the duality gap changes in each {\tt skydive} iteration. Here we see several climbing phases in order to avoid stalling. 
 
\begin{figure}[!ht]
	\centering
	\includegraphics[width=0.9\textwidth]{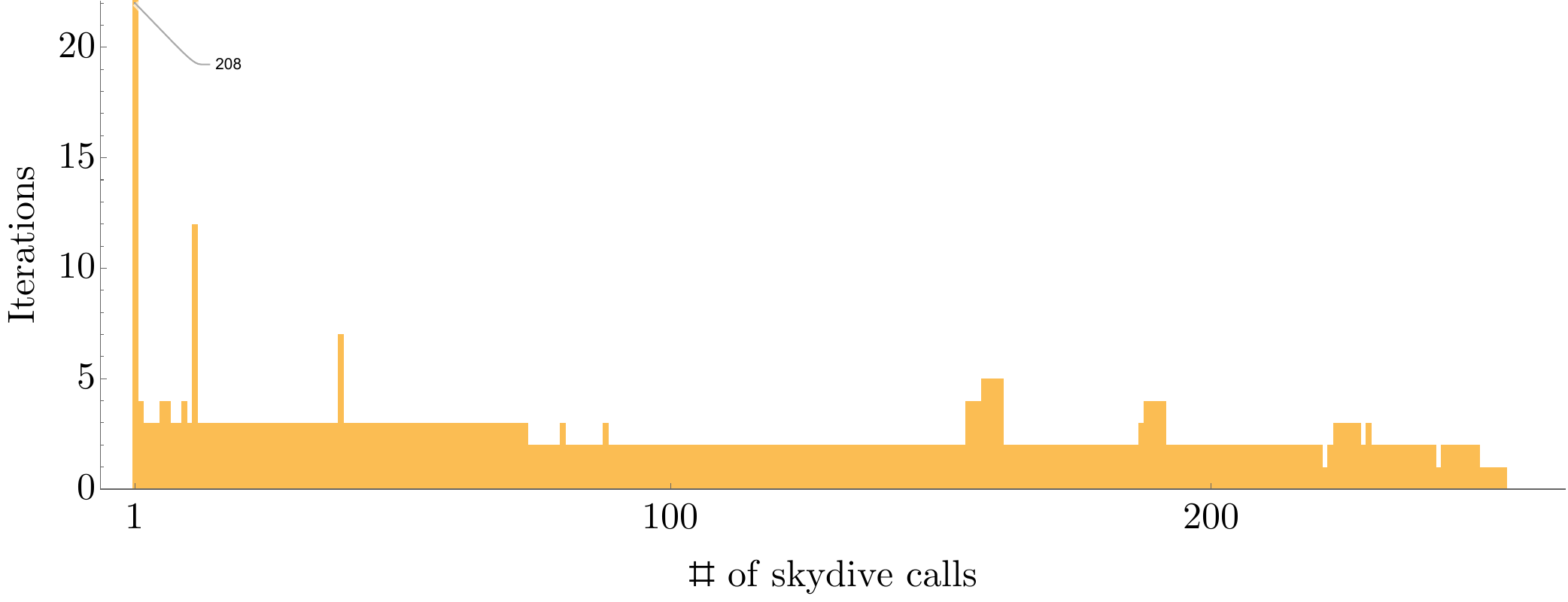}
	\caption[]{\label{fig:iters_vs_Npoint}The number of Newton iterations spent in each {\tt skydive call} for the skydiving run 1.}
\end{figure}

In figure~\ref{fig:iters_vs_Npoint}, the number of updates $d\xi$ conducted in each {\tt skydive} call is shown. These iterations include the pure PDIP steps in the initial  {\tt skydive} call  to reduce $\mu < \texttt{dualityGapUpperLimit} $, the centering steps which are PDIPs at $\beta=1$, and the full step updating both $(\xi, p)$. As we stressed before, these were computationally the most costly steps because of the required Cholesky factorization to invert the Hessian matrix $H_{\xi\xi}$ as  mentioned in section~\ref{subsec:review}. Once again we managed to realize a drastic reduction in the number of such iterations, thereby mitigating the bottleneck in standard \texttt{SDPB} computations. We also observe the typical behaviors of skydiving runs: the algorithm spends significantly many PDIP steps in the first {\tt skydive} call (208 in this example), and then the algorithm only requires a few iterations per call. 

We can compare results above with the (navigator $+$ SDPB) method. A navigator computation of the same problem was performed starting from:
$$
(0.51902209, 1.59623576, 1.20998, 2.988, 3.04429644, 2.42004559, 3.98537291,\\ 0.55365625).
$$
which is in between $p_1$ and $p_{\text final}$. This computation took 227 navigator function calls (so steps $dp$), which is again entirely in the same ballpark as the number of \texttt{skydive} calls. However, here one would normally require 240 PDIP (Newton) iterations to update $\xi$ in between updates of $p$. We therefore once again find a significant speedup. 
 
As an aside we note that we managed to find some improvement in the standard navigator method through the use of warmstarting: at a point $p_k$ we save a checkpoint about midway through the \texttt{SDPB} run, and then at point $p_{k+1}$ we restart \texttt{SDPB} from this checkpoint rather than completely un-initialized.  Normally such warmstarting would quickly result in stalling, but we managed to find a climbing routine (inspired by the scanning subroutine of {\tt skydive}, in particular item \ref{item:ifalphaisstilltoosmall} on page \pageref{item:ifalphaisstilltoosmall}, to recover from such stalling. After implementing the required modifications in \texttt{SDPB}\footnote{\url{https://github.com/suning-git/sdpb/tree/sdpb2.4.0_midck_stallingrecover}.} we were able to reduce the number of PDIP steps from 240 to about 60 on average.

We also compared the performance of the skydiving algorithm with the \texttt{tiptop} algorithm of \cite{Chester:2021}. In a specific \texttt{tiptop} run at $\Lambda=19$ \footnote{This computation was done during the project of \cite{Chester:2021}, but was not reported in the paper.}, the total number of Newton steps is about 143,910 \footnote{The \texttt{tiptop} algorithm scanned about 1,755 points in the space of $(\Delta_v,\Delta_s,\Delta_t,\Delta_{t_4})$. Feasibility at each point is determined by the cutting surface algorithm. On average, the cutting surface algorithm scans 41 points in the space of OPE coefficients, and each point corresponds to an SDP in feasibility mode, whose computation requires about 2 Newtonian steps.}. The skydiving run from $p_1$ has the number of iterations that are 173 times smaller! Also, the result from the skydiving run is almost exactly on the highest feasible value of $\Delta_{t_4}$, whereas the tiptop algorithm bisects the $\Delta_{t_4}$ until reaching a certain resolution.

\subsection{Usage tips and current limitations}\label{section:tips}
As we have alluded in previous sections, some of our parameters, \texttt{dualityGapUpperLimit},  \texttt{centeringRThreshold}, and the initial hessian $H_\text{init}$ to start the BFGS algorithm, affect the efficiency and the rate of success of skydiving runs and their preferred values depend closely on specific bootstrap problems and initial conditions. In this section, we will elaborate more on their roles during a run and offer some tips on how to choose them. We will also mention some current limitations of the algorithm

\subsubsection{Running off in parameter space}

Let us first discuss the importance of \texttt{dualityGapUpperLimit} and how it prevents the algorithm from leaving the region of interest in parameter space. We will use \textit{duality gap} and $\mu$ interchangeably in this section as they are proportional to each other according to \eqref {eq:duality gap mu}.  

As we mentioned in the beginning of section~\ref{section:Skydiving_Algorithm}, when the duality gap (or $\mu$) is large, the Lagrangian is not a reliable approximation to the navigator function, and we must avoid taking steps in parameter space until the duality gap falls below \texttt{dualityGapUpperLimit}. However, during the scanning steps discussed in section~\ref{section:scan}, it is possible for the duality gap to increase again, and potentially move back above \texttt{dualityGapUpperLimit}. Should we continue to impose \texttt{dualityGapUpperLimit} during scanning to prevent this?  

The parameter space of a bootstrap problem can contain multiple isolated feasible regions --- either compact regions (islands), or unbounded regions (continents). In a feasible region, the navigator function $\mathcal N(p)$ is non-positive. An island will contain at least one local minimum of $\mathcal N(p)$. The existence of multiple isolated feasible regions implies the non-convexity of the navigator function. Often, we want to restrict our attention to an island as we did in the previous subsections, but the non-convexity implies that the steps in parameter space $\cP$ can run off to a different region. The navigator method can mitigate this issue by penalizing the $\mathcal N(p) >0$ region.

By contrast, the finite-$\mu$ Lagrangian makes running-off behaviors more troublesome for a few reasons:
\begin{enumerate}
\item The finite-$\mu$ navigator generally starts large and gradually decreases. Unlike for the navigator function, there is no simple criterion like $\mathcal N(p)\geq 0$ for penalizing an unwanted region.
\item At large $\mu$, the Lagrangian is smoother and barriers between different local regions are lower, so it is easier for the solver to walk from one region to another.
\item If the finite-$\mu$ Lagrangian is too different from the navigator function, the solver can be attracted to a region that is different from the target region at $\mu=0$, and it may not be able to escape when $\mu$ gets smaller.
\end{enumerate}
We found that continuing to impose \texttt{dualityGapUpperLimit} can help avoid these complications.

If the initial point is sufficiently close to the island, we observed that a good choice of \texttt{dualityGapUpperLimit} for the first {\tt skydive} call is the value of the duality gap when the relative gap is $\bigO(1)$\footnote{Notice that this choice is problem dependent that might require users to launch some test run to find a  good \texttt{dualityGapUpperLimit}}:
\begin{equation}
\frac{|\text{primal objective}-\text{dual objective}|}{\text{min}(|\text{primal objective}|,|\text{dual objective}|)} \sim O(1),
\end{equation}

In the second and third run in table~\ref{tab:O3tiptopruns}, for example, we found that it is necessary to impose \texttt{dualityGapUpperLimit} on all {\tt skydive} iterations. The initial points $p_2$ and $p_3$ are significantly far from the island and a constant \texttt{dualityGapUpperLimit} prevents the solver from moving towards the ``continent". 

Another subtlety is that sometimes using $N_{pp}^\textrm{default}$ of \eqref{equ:Hmixed} as the initial Hessian can be problematic. In the Ising example, we observed that for the run from $p_1$, if we use $N_{pp}^\textrm{default}$ as the initial Hessian, the solver makes a bad step towards the continent, even though the gradient points towards the island. Most likely the problem is caused by off-diagonal elements in $N_{pp}^\textrm{default}$. We solved this problem by using the initial Hessian \eqref{equ:Ising_Hinit}, which is the diagonal piece of $N_{pp}^\textrm{default}$ at $p_1$.

\subsubsection{Failure of centering}
Our algorithm~\ref{alg:skydiving} assumes that the centering procedure in section~\ref{section:Skydiving_Algorithm} can bring $\xi_\mu(p)$ sufficiently close to the central section $\xi^*_\mu(p)$. In practice, this means that we assume that we are able to reduce $R_{err} \colonequals ||R||_{\text{max}}$ to arbitrarily small values at any fixed finite $\mu$.

However, in certain semidefinite programs it is not possible to find an optimal solution at finite $\mu$. One important class of examples is when primal or dual feasible points lie exactly on the boundary of the cone of positive semidefinite matrices rather than in its interior. This is a violation of the so-called Slater condition, which demands the existence of at least one feasible point which is strictly positive definite. The logarithmic barrier function introduced above is then infinite at every feasible point and consequently the central path does not really exist. The standard primal-dual interior point algorithm is then no longer guaranteed to work, although it might still perform well enough in practice.

In bootstrap studies, violation of the Slater condition happens occasionally on the dual side, when some components of the functionals must be exactly zero in order to satisfy the positivity conditions. In the course of an \texttt{SDPB} run, this typically corresponds to the failure to obtain dual feasible ``jumps;" instead the dual error reduces gradually to $0$ as $Y$ becomes increasingly singular.

Returning to the context of centering steps,  $R_{err}$ is a measure of how correctly the constraint $XY = \mu I$ is being solved and it is impossible to make $R_{err}$ zero at finite $\mu$ if there is no non-singular $Y$ satisfying the dual constraints. Hence, failing to satisfy the Slater condition manifests itself as a non-convergent $R_{err}$ in our algorithm. 

The current skydiving algorithm does not include a solution to this problematic situation. To avoid it, the user may have to modify their SDP, perhaps by discarding crossing equations by hand so that the Slater condition is satisfied. Another option is to bypass centering steps by setting \texttt{centeringRThreshold} to a large value. Finding a general strategy to cope with this scenario remains a problem for future research. 

\section{Conclusions and future directions}\label{sec:conclusion}

The core idea of skydiving is simple: combine optimization of a SDP with optimization over the external parameters that it depends on. In theory, this simply requires us to treat the Lagrange function of the SDP as a function of both internal variables $\xi$ and external parameters $p$. We have found that this idea can work in practice, as long as we supplement it with a few crucial techniques to avoid stalling: (1) centering iterations to stay close to the central section, (2) $\beta$-scan to carefully decide how to change $\mu$, (3) climbing in $\mu$ to recover from stalling when necessary. Furthermore, we found an efficient method for avoiding expensive computations of the Hessian in $p$-space using a BFGS-type update. These techniques are implemented in the open-source C++ program \texttt{skydive}.

Our tests of \texttt{skydive} on various conformal bootstrap problems show promising results. In some cases, the time spent towards solving an SDP is shorter than the time spent generating the SDP (by constructing matrices of conformal blocks and writing them to disk). In these cases, the bottleneck of the bootstrap computation is no longer in optimization and in the future we need to pursue other avenues for improving performance.

While \texttt{skydive} has already proved practically useful, it represents only an initial attempt in the direction of combined internal-external optimization. There are several possibilities for further enhancements, some of which we outline below.

The idea of iterating the corrector step has proven valuable to improve convergence at small computational cost. It would be interesting to use this idea in ordinary SDP solutions, as well as to use it to solve \eqref{eq:newton-minimize} with fixed $dp$.\footnote{In our \texttt{skydive} program, we have actually implemented corrector iterations for solving \eqref{eq:newton-minimize} with fixed $dp$. However, this requires saving the Schur complement from the previous SDP and loading it into the new SDP. In the current implementation, this data is processed as text, resulting in slow read/write operations. We find it to be inefficient in practice, but it can certainly be improved in the future.} We expect that this could make it relatively cheap to move along the local central path.

In skydiving, there is a basic dilemma: at higher values of $\mu$, the solver can take larger steps in $d\xi$ (since the $X,Y$ matrices are not close to degeneracy), but the prediction for $p$ becomes less accurate (due to the large-$\mu$ navigator function being a poor approximation to the true $\mu=0$ navigator function). Meanwhile, the opposite holds for lower values of $\mu$. In our current algorithm, we seek a balance between these competing effects using the ``$\beta$-scan" and ``climbing" techniques. However, if it becomes relatively inexpensive to move along the local central path, it might be appealing to implement a strategy where the solver computes the step $dp$ at lower $\mu$ but executes the step at higher $\mu$. A natural measure of the accuracy of the navigator function could be the relative duality gap 
\be
\frac{|\text{primal objective}-\text{dual objective}|}{\min(|\text{primal objective}|,|\text{dual objective}|)},
\ee
 while a natural criterion for determining whether $\mu$ is large enough could come from testing the step length.

Our current implementation separates control of the main loop and the computations inside the loop between the external ``driver" programs and \texttt{skydive}. While this implementation is suitable for experimentation, it suffers from high I/O costs, due to the need to repeatedly generate new SDP files and write them to disk in the driver, and load them from disk in \texttt{skydive}. In the future, it would be desirable to develop an integrated program that executes the full algorithm. 

A more ambitious goal is to extrapolate bootstrap computations from lower to higher derivative order.\footnote{This idea was made to work for the spinless modular bootstrap in \cite{Afkhami-Jeddi:2019zci}.} As the derivative order increases, the data specifying the SDP ($b$, $B$, and $c$) will increase in dimension discontinuously. An important challenge will be to compute an appropriate $d\xi$ to account for this discontinuous change.

\section*{Acknowledgments}

We thank Brian McPeak, Matt Mitchell, Marten Reehorst, Slava Rychkov, Benoit Sirois, and Alessandro Vichi for discussions. We would also like to thank the organizers of the `Positivity' workshop at the PCTS for providing a stimulating environment. AL and DSD are supported by Simons Foundation grant no.\ 488657 (Simons Collaboration on the Nonperturbative Bootstrap) and a DOE Early Career Award under grant no.\ DE-SC0019085. This project received funding from the European Research Council (ERC) under the European Union Horizon 2020 research and innovation programme (grant agreement no.\ 758903). BvR is supported by Simons Foundation grant no.\ 488659 (Simons Collaboration on the Nonperturbative bootstrap). The computations in this work were run on the Symmetry cluster at the Perimeter institute and the Resnick High Performance Computing Center, a facility supported by Resnick Sustainability Institute at the California Institute of Technology. AL and NS thank the Perimeter Institute for hosting the Mini-Course of Numerical Conformal Bootstrap in April 2023, which included tutorials on skydiving and associated software tools. Research at Perimeter Institute is supported in part by the Government of Canada through the Department of Innovation, Science and Economic Development Canada and by the Province of Ontario through the Ministry of Colleges and Universities.

\bibliographystyle{utphys}
\bibliography{skydiving_draft}

\providecommand{\href}[2]{#2}\begingroup\raggedright\begin{thebibliography}{10}

\bibitem{Poland_2019}
D.~Poland, S.~Rychkov, and A.~Vichi, ``The conformal bootstrap: Theory,
  numerical techniques, and applications,''
  \href{http://dx.doi.org/10.1103/revmodphys.91.015002}{{\em Reviews of Modern
  Physics} {\bfseries 91} no.~1, (Jan, 2019) }.
  \url{https://doi.org/10.1103%2Frevmodphys.91.015002}.

\bibitem{poland2022snowmass}
D.~Poland and D.~Simmons-Duffin, ``Snowmass white paper: The numerical
  conformal bootstrap,'' 2022.

\bibitem{Reehorst:2021ykw}
M.~Reehorst, S.~Rychkov, D.~Simmons-Duffin, B.~Sirois, N.~Su, and B.~van Rees,
  ``{Navigator Function for the Conformal Bootstrap},''
  \href{http://dx.doi.org/10.21468/SciPostPhys.11.3.072}{{\em SciPost Phys.}
  {\bfseries 11} (2021) 072}, \href{http://arxiv.org/abs/2104.09518}{{\ttfamily
  arXiv:2104.09518 [hep-th]}}.

\bibitem{Poland:2011ey}
D.~Poland, D.~Simmons-Duffin, and A.~Vichi, ``{Carving Out the Space of 4D
  CFTs},'' \href{http://dx.doi.org/10.1007/JHEP05(2012)110}{{\em JHEP}
  {\bfseries 05} (2012) 110}, \href{http://arxiv.org/abs/1109.5176}{{\ttfamily
  arXiv:1109.5176 [hep-th]}}.

\bibitem{Kos:2014bka}
F.~Kos, D.~Poland, and D.~Simmons-Duffin, ``{Bootstrapping Mixed Correlators in
  the 3D Ising Model},'' \href{http://dx.doi.org/10.1007/JHEP11(2014)109}{{\em
  JHEP} {\bfseries 11} (2014) 109},
  \href{http://arxiv.org/abs/1406.4858}{{\ttfamily arXiv:1406.4858 [hep-th]}}.

\bibitem{Simmons-Duffin:2015qma}
D.~Simmons-Duffin, ``{A Semidefinite Program Solver for the Conformal
  Bootstrap},'' \href{http://dx.doi.org/10.1007/JHEP06(2015)174}{{\em JHEP}
  {\bfseries 06} (2015) 174}, \href{http://arxiv.org/abs/1502.02033}{{\ttfamily
  arXiv:1502.02033 [hep-th]}}.

\bibitem{Landry:2019qug}
W.~Landry and D.~Simmons-Duffin, ``{Scaling the semidefinite program solver
  SDPB},'' \href{http://arxiv.org/abs/1909.09745}{{\ttfamily arXiv:1909.09745
  [hep-th]}}.

\bibitem{doi:10.1137/0802028}
S.~Mehrotra, ``On the implementation of a primal-dual interior point method,''
  \href{http://dx.doi.org/10.1137/0802028}{{\em SIAM Journal on Optimization}
  {\bfseries 2} no.~4, (1992) 575--601},
  \href{http://arxiv.org/abs/https://doi.org/10.1137/0802028}{{\ttfamily
  https://doi.org/10.1137/0802028}}. \url{https://doi.org/10.1137/0802028}.

\bibitem{yamashita2003implementation}
M.~Yamashita, K.~Fujisawa, and M.~Kojima, ``Implementation and evaluation of
  sdpa 6.0 (semidefinite programming algorithm 6.0),'' {\em Optimization
  Methods and Software} {\bfseries 18} no.~4, (2003) 491--505.

\bibitem{yamashita2010high}
M.~Yamashita, K.~Fujisawa, K.~Nakata, M.~Nakata, M.~Fukuda, K.~Kobayashi, and
  K.~Goto, ``A high-performance software package for semidefinite programs:
  Sdpa 7,'' {\em Tokyo, Japan} (2010) .

\bibitem{yamashita2012latest}
M.~Yamashita, K.~Fujisawa, M.~Fukuda, K.~Kobayashi, K.~Nakata, and M.~Nakata,
  ``Latest developments in the sdpa family for solving large-scale sdps,'' {\em
  Handbook on semidefinite, conic and polynomial optimization} (2012) 687--713.

\bibitem{shapiro1997first}
A.~Shapiro, ``First and second order analysis of nonlinear semidefinite
  programs,'' {\em Mathematical programming} {\bfseries 77} (1997) 301--320.

\bibitem{leibfritz2002interior}
F.~Leibfritz and E.~Mostafa, ``An interior point constrained trust region
  method for a special class of nonlinear semidefinite programming problems,''
  {\em SIAM Journal on Optimization} {\bfseries 12} no.~4, (2002) 1048--1074.

\bibitem{correa2004global}
R.~Correa, ``A global algorithm for nonlinear semidefinite programming,'' {\em
  SIAM Journal on optimization} {\bfseries 15} no.~1, (2004) 303--318.

\bibitem{kovcvara2003pennon}
M.~Ko{\v{c}}vara and M.~Stingl, ``Pennon: A code for convex nonlinear and
  semidefinite programming,'' {\em Optimization methods and software}
  {\bfseries 18} no.~3, (2003) 317--333.

\bibitem{yamashita2015survey}
H.~Yamashita and H.~Yabe, ``A survey of numerical methods for nonlinear
  semidefinite programming,'' {\em Journal of the Operations Research Society
  of Japan} {\bfseries 58} no.~1, (2015) 24--60.

\bibitem{enwiki:1150228229}
{Wikipedia contributors}, ``Broyden–fletcher–goldfarb–shanno algorithm
  --- {Wikipedia}{,} the free encyclopedia,'' 2023.
\newblock
  \url{https://en.wikipedia.org/w/index.php?title=Broyden%E2%80%93Fletcher%E2%80%93Goldfarb%E2%80%93Shanno_algorithm&oldid=1150228229}.
  [Online; accessed 11-June-2023].

\bibitem{Kos:2015mba}
F.~Kos, D.~Poland, D.~Simmons-Duffin, and A.~Vichi, ``{Bootstrapping the O(N)
  Archipelago},'' \href{http://dx.doi.org/10.1007/JHEP11(2015)106}{{\em JHEP}
  {\bfseries 11} (2015) 106}, \href{http://arxiv.org/abs/1504.07997}{{\ttfamily
  arXiv:1504.07997 [hep-th]}}.

\bibitem{Su:2022xnj}
N.~Su, ``{The Hybrid Bootstrap},''
  \href{http://arxiv.org/abs/2202.07607}{{\ttfamily arXiv:2202.07607
  [hep-th]}}.

\bibitem{Chester:2021}
S.~M. Chester, W.~Landry, J.~Liu, D.~Poland, D.~Simmons-Duffin, N.~Su, and
  A.~Vichi, ``Bootstrapping heisenberg magnets and their cubic instability,''
  \href{http://dx.doi.org/10.1103/PhysRevD.104.105013}{{\em Phys. Rev. D}
  {\bfseries 104} (Nov, 2021) 105013}.
  \url{https://link.aps.org/doi/10.1103/PhysRevD.104.105013}.

\bibitem{Go:2019lke}
M.~Go and Y.~Tachikawa, ``{autoboot: A generator of bootstrap equations with
  global symmetry},'' \href{http://dx.doi.org/10.1007/JHEP06(2019)084}{{\em
  JHEP} {\bfseries 06} (2019) 084},
  \href{http://arxiv.org/abs/1903.10522}{{\ttfamily arXiv:1903.10522
  [hep-th]}}.

\bibitem{Afkhami-Jeddi:2019zci}
N.~Afkhami-Jeddi, T.~Hartman, and A.~Tajdini, ``{Fast Conformal Bootstrap and
  Constraints on 3d Gravity},''
  \href{http://dx.doi.org/10.1007/JHEP05(2019)087}{{\em JHEP} {\bfseries 05}
  (2019) 087}, \href{http://arxiv.org/abs/1903.06272}{{\ttfamily
  arXiv:1903.06272 [hep-th]}}.

\end{thebibliography}\endgroup

\appendix

\end{document}